\documentclass[ba]{imsart}

\RequirePackage{amsthm,amsmath,amsfonts,amssymb}
\RequirePackage[authoryear]{natbib}%% uncomment this for author-year citations
\RequirePackage[colorlinks,citecolor=blue,urlcolor=blue,backref=page,backref=page]{hyperref}
\RequirePackage{graphicx}

\RequirePackage{bm}
\RequirePackage{ams math}
\RequirePackage{color}

\RequirePackage{adjustbox}
\RequirePackage[graphicx]{realboxes}
\RequirePackage{rotating}

\RequirePackage{multicol}  %table
\RequirePackage{multirow} %table
\RequirePackage{booktabs}     %table
\RequirePackage{threeparttable}      %table
\RequirePackage{float}

\RequirePackage[ruled,linesnumbered]{algorithm2e}

\newcommand{\T}{\!\top\!}
\def\bfm#1{\mbox{\boldmath$#1$}}

\newcommand{\bI}{{\bf I}}

\newcommand{\bbT}{{\mathbb{T}}}
\newcommand{\bbI}{{\mathbb{II}}}
\newcommand{\bbR}{{\mathbb{R}}}
\newcommand{\bbU}{{\mathbb{U}}}
\newcommand{\bbV}{{\mathbb{V}}}
\newcommand{\bbZ}{{\mathbb{Z}}}
\newcommand{\bbD}{{\mathbb{D}}}
\newcommand{\bbX}{{\mathbb{X}}}
\newcommand{\bx}{{\bf x}}
\newcommand{\ba}{{\bf a}}

\newcommand{\by}{{\bm y}}
\newcommand{\bX}{{\bf X}}
\newcommand{\bY}{{\bf Y}}

\newcommand{\bu}{{\bf u}}
\newcommand{\bU}{{\bf U}}
\newcommand{\bW}{{\bf W}}

\newcommand{\bthe}{{\bfm \theta}}
\newcommand{\bmX}{{\boldsymbol{\mathcal{X}} }}
\newcommand{\bmY}{{\boldsymbol{\mathcal{Y}} }}
\newcommand{\bmW}{{\boldsymbol{\mathcal{W}} }}
\newcommand{\bmF}{{{\mathcal{F}} }}
\newcommand{\0}{{\bf 0}\!\!\!{\bf 0}}
\newcommand{\bSi}{{\bfm \Sigma}}
\newcommand{\bOm}{{\bfm \Omega}}
\newcommand{\bsi}{{\bfm \sigma}}
\newcommand{\bPhi}{{\bfm \Phi}}

\newcommand{\bva}{{\bfm \varphi}}
\newcommand{\bla}{{\bfm \lambda}}
\newcommand{\bmu}{{\bfm \mu}}
\newcommand{\bxi}{{\bfm \xi}}
\newcommand{\bal}{{\bfm \alpha}}
\newcommand{\E}{{\rm E}}

\newcommand{\bpsi}{{\bfm \psi}}
\pubyear{2024}
\arxiv{2010.00000}
\volume{TBA}
\issue{TBA}
\firstpage{1}
\lastpage{1}

\startlocaldefs
\theoremstyle{plain}

\theoremstyle{definition}

\theoremstyle{remark}

\endlocaldefs

\begin{document}
	
	\begin{frontmatter}
		\title{Variational Bayesian Logistic Tensor Regression with Application to Image Recognition}
		%\title{A sample article title with some additional note\thanksref{t1}}
		\runtitle{VBLTR with Application to Image Recognition}
		%\thankstext{T1}{A sample additional note to the title.}
		
		\begin{aug}
			%%%%%%%%%%%%%%%%%%%%%%%%%%%%%%%%%%%%%%%%%%%%%%%
			%% Only one address is permitted per author. %%
			%% Only division, organization and e-mail is %%
			%% included in the address.                  %%
			%% Additional information can be included in %%
			
			\author[A]{\fnms{Yunzhi}~\snm{Jin}\ead[label=e1]{yzjin@ynu.edu.cn}},
			\author[B]{\fnms{Yanqing}~\snm{Zhang}\ead[label=e2]{zyqznl2010@126.com}}
			\and
			\author[C]{\fnms{Niansheng}~\snm{Tang}\ead[label=e3]{nstang@ynu.edu.cn}\orcid{0000-0001-7033-3845}}
			%%%%%%%%%%%%%%%%%%%%%%%%%%%%%%%%%%%%%%%%%%%%%%
			%% Addresses                                %%
			%%%%%%%%%%%%%%%%%%%%%%%%%%%%%%%%%%%%%%%%%%%%%%
			\address[A]{Yunnan Key Laboratory of Statistical Modeling and Data Analysis,
				Yunnan University\printead[presep={,\ }]{e1}}
			
			\address[B]{Yunnan Key Laboratory of Statistical Modeling and Data Analysis,
				Yunnan University\printead[presep={,\ }]{e2}}
			
			\address[C]{Yunnan Key Laboratory of Statistical Modeling and Data Analysis,
				Yunnan University\printead[presep={,\ }]{e3}}
			
			\runauthor{Y. Jin, Y. Zhang, and N. Tang}
		\end{aug}
		
		\begin{abstract}
			
			In recent years, image recognition method has been a research hotspot in various fields such as video surveillance, biometric identification, unmanned vehicles, human-computer interaction, and medical image recognition. Existing recognition methods often ignore structural information of image data or depend heavily on the sample size of image data.
			To address this issue, we develop a novel variational Bayesian method for image classification
			in a logistic tensor regression model with image tensor predictors by utilizing tensor decomposition to approximate tensor regression.
			To handle the sparsity of tensor coefficients, we introduce the multiway shrinkage priors for marginal factor vectors of tensor coefficients.
			In particular, we obtain a closed-form approximation to the variational posteriors for classification prediction based on the matricization of tensor decomposition. Simulation studies are conducted to investigate the performance of the proposed methodologies in terms of accuracy, precision and F1 score.  Flower image data and chest X-ray image data are illustrated by the proposed methodologies.
			
		\end{abstract}
		
		
		
	\end{frontmatter}
	
	\section{Introduction}
	
	Image recognition or classification is an important research topic and a widely used computer vision-based machine learning technique in various fields including video surveillance, biometric identification, unmanned vehicles, human-computer interaction, and medical image recognition. Many methods have been developed for image recognition such as K-nearest neighbors algorithm (KNN), random forests (RF), C-support vector classification (C-SVC) \citep{Boser1992,Cortes1995}, $\nu$-support vector regression ($\nu$-SVR) \citep{Scholkopf2000}, linear-support vector regression (L-SVR) \citep{Ho2012}, among others. The aforementioned methods mainly focus on
	the vector or matrix reconstituted or extracted from image data. In many practical applications, image data saved in a computer may be the three-order tensor data, which are multi-dimensional arrays and retain much underlying structural information and feature information.
	Although tensor vectorization or matricization is an invertible map and can preserve the correlation structure of tensor data,
	vectorizing or matricizing tensor data may destroy the inherent spatial structure in the original data %that possesses wealth of information
	when tensor data has a more complicated structure \citep{Zhou2013}.
	Thus, the vector or matrix reconstituted or extracted from image data
	may lose a wealth of spatial structural information and important features of images. More importantly, the vectorization of a tensor may lead to a potentially high dimensional data, easily yielding overtraining, highly computational complexity and large memory requirement.
	
	In recent years, some tensor representation based techniques have been proposed to address the aforementioned issues.
	For example, \cite{4429309} introduced the correlation concept to tensor representations, and presented a correlation tensor analysis for constructing a maximal correlation affinity classifier.
	\cite{7389377} employed the multi-linear principal component analysis to preprocess tensors, and developed support tensor machine for image classification. \cite{8066324} presented a new tensor-based feature extraction algorithm, called tensor rank preserving discriminant
	analysis, for facial image recognition. \cite{Zhao_2020_CVPR} developed the pairwise self-attention and patchwise self-attention methods for image recognition.
	\cite{Shi2022} proposed an end-to-end and pixel-to-pixel IoT-oriented fuzzy support tensor product adaptive method
	for image classification. Deep learning techniques have been developed for tensor image recognition on the basis of various variations of convolutional neural networks (CNN) \citep{8320684}.
	 Also,
		various tensor regressions have been proposed for image data analysis, including scalar-on-tensor regression which is a regression model with a scalar response and tensor-valued covariates \citep{Tan2013,Zhou2013,Ming2015,Masaaki2016,Li2018,Fang2019,Le2020,Esa2022,LU2024105249}, tensor-on-tensor regression which is a regression model with a tensor-valued response and a tensor-valued covariates \citep{peter2015,Lock2018,Liu2020,Lee2024}, and tensor-on-vector regression which is a regression model with a tensor-valued response and a vector-valued covariates \citep{Qi2020TensorToVectorRF}. In particular,
		\cite{Zhou2013} developed a generalized tensor regression and proposed a maximum likelihood estimation for tensor coefficients based on CANDECOMP/PARAFAC (CP) decomposition \citep{carroll70,Hars1970}.
	\cite{8408516} considered logistic tensor regression, studied its estimation based on rank-1 canonical decomposition, and developed a rank-1 feed-forward neural network for image analysis.
	\cite{8877994} studied a generalized tensor regression
	and presented the penalized least squared procedure to estimate tensor coefficients via the quadratic penalty.
	\cite{Konyar2024} proposed a federated generalized scalar-on-tensor
		regression framework where multiple local tensor models are learned at the edge, and their parameters are shared and
		updated by an aggregator.
		\cite{Zhou2024} proposed a nonparametric scalar-on-tensor regression and presented a low-rank estimation approach based on the
		Polynomial splines and elastic-net penalty function.
	Although the preceding introduced methods considered the structural information of image via tensor representation or tensor regression, they ignored prior knowledge on tensor modes or tensor decomposition,
	which may affect the accuracy of image recognition.
	
	%	\cite{Rajarshi2021,Sharmistha2021,Lee2022}
	
	Bayesian approaches can be used to overcome the aforementioned issues with some proper prior information of tensor modes or factors associated with tensor decomposition in the Markov chain Monte Carlo (MCMC) framework. For example, see \cite{peter2015} for tensor-on-tensor regression, \cite{JMLR:v18:16-362} for constructing the shrinkage priors of tensor coefficients in scale-on-tensor regression, \cite{Billio2023} for linear autoregressive tensor model, and \cite{Wu2022Tang} for tensor logistic regression,	\cite{Lei2024} for vector-on-tensor regression,
	\cite{Billio2024} for tensor-on-tensor regression model.
		Although the MCMC algorithm can utilize the prior information of tensor modes or factors,
		the Markov chains in implementing the MCMC algorithm require being carefully constructed so that
		their limiting distribution matches the target distribution, which is quite difficult or almost impossible for some complicated posterior distributions in many applications. Also, traditional MCMC approaches suffer from a number of drawbacks, including low sampling efficiency due to sample correlation and a lack of computational scalability to massive datasets \citep{ZhangYang2024}. Moreover, the MCMC algorithm requires huge memory and intensive computational cost in handling tensor data associated with image data.
	To this end, variational Bayesian approach \citep{beal2003}, which is an effective computational method to
		approximate the complicated posterior distributions in the Bayesian framework, 	
	may be employed to alleviate computational cost and memory requirement
	as an alternative to the MCMC algorithm.
	Due to its merits such as high-speed computation and without involving intractable multiple integrals, % and small memory required,
	variational Bayesian approach \citep{beal2003}
	has received considerable attention in recent years. For instance, see \cite{Wipf2011Rao} for latent variable models,
	\cite{Ghahramani2000Beal} for mixtures of factor analysis,
	\cite{Attias2000} for graphical models, \cite{Wu2021Tang} for partially linear mean shift models with high-dimensional data,
	\cite{Yi2022Tang} for high-dimensional linear mixed models, and \cite{NIPS2011_b495ce63,Li2013,9737135}
	for image recognition based on feature matrix extracted from image data,
	which may lose some important feature information.
	However, to our knowledge, there is little work on variational Bayesian inference for tensor regression models.

	To this end, we develop a novel variational Bayesian method to estimate tensor coefficients (denoted as VBLTR) in a tensor logistic regression model for image recognition with any tensor image predictors rather than the vectorization forms of tensor image predictors.
	Since tensor predictors can retain considerable correlation and structural information and the vectorization of tensor image data may corrupt the intrinsic structure of image data,
	logistic tensor regression can incorporate more structural information than the vector-based logistic model.
	To reduce the number of parameters, tensor coefficient in the considered logistic tensor regression is
		approximated with a low-rank structure via CANDECOMP/PARAFAC (CP) decomposition \citep{carroll70,Hars1970}.
	We utilize the matricization idea of tensor approximation via CP decomposition to deduce different equivalent expressions of logistic tensor regression approximation. Based on the approximated logistic tensor regression and the multiway shrinkage priors for marginal factor vectors of tensor coefficient approximation, we develop a variational Bayesian method to obtain the sparse estimator of tensor coefficient approximation. To implement variational Bayesian method, we introduce a variational parameter $\xi$ and employ $\xi$-transformation technique \citep{Jaakkola2000BayesianPE} to approximate the evidence lower bound, which needs less computational cost compared with the Polya-Gamma data augmentation scheme
		used in \cite{Wu2022Tang} in that the proposed method avoids Markov chain sampling due to non-randomness of $\xi$. Also, we present an approximation to predictive density based on the deduced variational posteriors for classified prediction.
	%Some simulation studies are conducted to assess the classification performance of the proposed method. We analyze flower image data %and chest X-ray image data and show that the proposed method for image classification outperforms other comparing methods.

	The proposed method has fifth significant contributions.
	First, the proposed method can retain the correlation and structural information of tensor data
	in improving the accuracy of image recognition.
	In practical applications such as image analysis, video analysis or recommendation systems,
	data often take the form of multidimensional arrays or tensors and possess the strong correlation and structure information.
	Traditional vector-based logistic regression commonly neglects the correlation and structural information.
	In comparison with vector-based approaches, the developed method introduces the underlying structural prior with lower dimensions and retains the intrinsic structural information of image data for efficiently improving classification accuracy.
	Second, a variational Bayesian approach is developed to estimate the approximated tensor coefficients via the CP decomposition of tensor coefficients, which can incorporate the prior information of the correlated and sparse structure and avoid the multiple integrals involved in image data analysis. Since variational Bayesian approach doesn't require huge memory and converges fast in contrast to the traditional MCMC approaches, the proposed method can obtain Bayesian estimation of tensor coefficients with the high-speed computation. To our knowledge, we are the first to develop a tensor estimation procedure for a tensor logistic regression in the variational Bayesian framework.
	Third, the proposed method utilizes the matricization idea of tensor coefficient approximation via CP decomposition, the one-order approximation of the predictive density, % and $\xi$-transformation,
	which can largely reduce the complexity of estimation and prediction.
	Since the matricization of the tensor coefficient approximation via CP decomposition can be expressed as different equivalent forms of the product of factor matrices, we can develop the profile estimators of each factor matrix based on different expressions of tensor coefficient approximation, which avoids complex matrix operations. The one-order approximation of the predictive density deduced from variational posteriors avoids intractable multiple integrals and complicated exponential function calculation, which makes classification prediction more efficient.
	Fourth, unlike \cite{Wu2022Tang}, the proposed method employs the multiway Dirichlet generalized double Pareto (M-DGDP) prior to induce the shrinkage components of parameters in CP decomposition of coefficient tensor,
		which can be used to identify sparse structures in CP decomposition of coefficient tensor. In particular, we propose
		using the $\xi$-transformation rather than the Polya-Gamma data augmentation scheme to approximate the evidence lower bound
		even if the latter can de-marginalize the binomial likelihood, remove the need to use the $g(\cdot)$ link function
		and allow for conjugate Gaussian prior. The reasons are (i)
		$\xi$-transformation also avoids the usage of logistic function $g(\cdot)$ link function with unknown coefficient parameters; (ii) the Polya-Gamma variable $\eta$ associated with the Polya-Gamma data augmentation scheme is random and needs to evaluate its optimal variational density, while
		variational parameter $\xi$ associated with the proposed method is not random and has a closed-form,
		which shows that the proposed method can avoid to estimate the optimal variational density of extra parameters and Markov chain sampling
		at each iteration; (iii) the Polya-Gamma data augmentation is implicitly hidden in the quadratic lower bound of the considered $\xi$-transformation \citep{Daniele2019}.
		Fifth, we conduct numerical comparison with the existing approach \citep{Wu2022Tang}
		for assessing the relative performance of the proposed method.

	The rest of this paper is organized as follows. Section 2 introduces the notation and some basic concepts on tensor. Section 3 develops a variational Bayesian estimation for tensor coefficients and predictive density in a logistic tensor regression.
	Simulation studies are conducted to assess the performance of the proposed approach in Section 4. In Section 5,
	we illustrate the usage of the proposed method with flower image data and chest X-ray image data.
	Concluding remarks and discussion are given in Section 6. Some technical details are presented in the Supplementary Material \citep{Jin2025}.

	\section{Notation and Preliminaries}
	
	\subsection{Tensor and Its Decomposition}
	
	{\bf Notation}. In this subsection, we introduce some notation and tensor's decomposition.
	Throughout this article, we use blackboard capital letters for sets,
	e.g., $\bbT,\bbI$, small letters for scalars, e.g., $x,y\in\bbR$,
	bold small letters for vectors, e.g., $\bx, \by\in\bbR^I$, bold capital letters for matrices, e.g., $\bX, \bY\in\bbR^{I_1\times I_2}$,
	and Euler script fonts for tensors (i.e., the number of their modes is strictly larger than two), e.g., $\bmX,\bmY\in\bbR^{I_1\times I_2\times\cdots\times I_M}\; (M>2)$.
	
	{\bf $M$-order Tensor}. A $M$th-order tensor is an array with $M$ dimensions ($M> 2$), which is an extension of a matrix to a higher order array, where $M$ represents the order of the tensor.
	We denote the $(i_1,i_2,\cdots, i_M)$th component of a $M$th-order tensor $\bmX\in\bbR^{I_1\times I_2\times\cdots\times I_M}$ as
	$x_{i_1i_2\cdots i_M}$, where $i_k=1,2,\ldots,I_k$, and $k$ is called a mode of the tensor for $k=1,2,\ldots,M$.
	In particular, a tensor $\bmX$ is called a rank-one tensor if it can be written as
	$\bmX=\bu^{(1)}\circ\bu^{(2)}\circ\cdots\circ\bu^{(M)}$,
	where the symbol $\circ$ represents the vector outer product, and
	$\bu^{(k)}=(u^{(k)}_1,u^{(k)}_2,\cdots,u^{(k)}_{I_k})^{\T}$ is a $I_k$-dimensional latent factor corresponding to the $k$th mode.
	Thus, each component of tensor $\bmX$ has the form: $x_{i_1i_2\cdots i_M}=u^{(1)}_{i_1}u^{(2)}_{i_2}\cdots u^{(M)}_{i_M}$.

	{\bf Matricization (Mode-$d$ unfolding)}. Reordering the components of tensor $\bmX$, we can unfold
	$\bmX$ into a matrix, which is referred to as matricization \citep{carroll70,Hars1970}.
	The mode-$d$ matricization of $\boldsymbol{\mathcal{X}}$, denoted as $\mathbf{X}_{(d)} \in \mathbb{R}^{I_{d} \times\left(I_{1} \ldots I_{d-1} I_{d+1} \ldots I_{M}\right)}$, is a procedure of mapping the components from a multidimensional array to a matrix, which is implemented by stacking mode-$d$ fibers (modal vectors) as column vectors of the resulting matrix, where mode-$d$ fiber of $\bmX$ is a vector of tensor components obtained by fixing all but one index of $\bmX$ (e.g., $\mathbf{x}_{i1,i2,\cdots,i_{d-1},:,i_{d+1},\cdots,i_M}\in \bbR^{I_d}$). Thus, the $(i_1,i_2,\ldots,i_M)$th component of $\bmX$ maps to the $(i_d, j)$th component of $\mathbf{X}_{(d)}$, where
	$j=1+\sum_{k=1, k \neq d}^{M} \{(i_{k}-1) %J_{k} \quad \textmd{with} \quad J_{k}=
	\prod_{m=1, m \neq d}^{k-1} I_{m}\}$.

	{\bf Tensor product}. The inner product of two tensors $\boldsymbol{\mathcal{X}}, \boldsymbol{\mathcal{Y}} \in \mathbb{R}^{I_{1} \times I_{2} \cdots \times I_{M}}$ with the same size is defined as the sum of the products of their corresponding elements:
	$$
	\langle\boldsymbol{\mathcal{X}}, \boldsymbol{\mathcal{Y}}\rangle=\sum_{i_{1}=1}^{I_{1}} \cdots \sum_{i_{M}=1}^{I_{M}} x_{i_{1} \ldots i_{M}} y_{i_{1} \ldots i_{M}},
	$$
	which can be rewritten as
	\begin{equation}\label{unfolded}
		\langle\boldsymbol{\mathcal{X}}, \boldsymbol{\mathcal{Y}}\rangle=\langle\mathbf{X}_{(j)}, \mathbf{Y}_{(j)}\rangle={\rm trace}(\mathbf{X}_{(j)} \mathbf{Y}_{(j)}^{\top})\;\;{\rm for}\;\; j=1,2,\ldots,M.
	\end{equation}

	{\bf Tensor decomposition}. The CANDECOMP/PARAFAC (CP) decomposition \citep{carroll70,Hars1970} of tensor $\boldsymbol{\mathcal{X}} \in \bbR^{I_{1} \times I_{2} \cdots \times I_{M}}$ is defined as a sum of $R$ rank-one tensors:
	\begin{equation}\label{CP}
		\bmX \approx \sum_{r=1}^{R} \bu_{\chi r}^{(1)} \circ \bu_{\chi r}^{(2)} \cdots \circ \bu_{\chi r}^{(M)}. % \triangleq [[\bU^{(1)}, \bU^{(2)}, \ldots, \bU^{(M)} ]],
	\end{equation}
	Denote $\bU_\chi^{(k)}=[\bu_{\chi 1}^{(k)}, \ldots, \bu_{\chi R}^{(k)}] \in \bbR^{I_{k} \times R}$ as factor matrices for $k=1,2, \ldots, M$.
	When $R$ is equal to the rank of tensor $\bmX$, the above defined CP decomposition (\ref{CP}) is equal to $\bmX$.
	The rank of tensor $\bmX$ is defined as the smallest number of rank-one tensors whose sum is equal to $\bmX$.
%	Figure \ref{realdata2fig01} displays the CP decomposition of a 3rd-order tensor \citep{carroll70,Hars1970}.
	Based on the CP decomposition of $\bmX\in \bbR^{I_{1} \times I_{2} \cdots \times I_{M}}$, the mode-$d$ matricization of $\bmX$
	can be approximated by
	$$
	\bX_{(d)}\approx\bU_\chi^{(d)}\left(\bU_\chi^{(M)} \odot \cdots \odot \bU_\chi^{(d+1)} \odot \bU_\chi^{(d-1)} \odot \cdots \odot \bU_\chi^{(1)}\right)^{\top}
	$$
	for $d=1,2,\ldots,M$, where the symbol $\odot$ represents the Khatri-Rao product of two matrices.

%		\begin{figure}[H]
%		\centering
%		\scalebox{1.2}[1.2]{\includegraphics{CPD.png}}\vspace{-3mm}
%		\caption{\footnotesize{CP decomposition of a 3rd-order tensor}}
%		\label{realdata2fig01}
%	\end{figure}

	\subsection{Variational Bayesian Inference}
	
	In this subsection, we introduce the basic idea of variational Bayesian inference in a parametric model $\mathcal{M}$. To this end, we consider the
	dataset $\bbD=\{\bbX,\bfm{y}\}$ associated with $\mathcal{M}$,
	where $\bbX$ is the set of covariates and $\bfm{y}$ is the vector of observations for response variables.
	Suppose that $\bfm{y}$ has the joint probability density function $p(\bfm{y}|\bbX,\bbV,\bpsi)$,
	where $\bbV$ is the set of latent variables, and $\bpsi$ is the set of parameters in the considered model $\mathcal{M}$. Here we
	assume that $\bbV$ has the probability density function $p(\bbV|\bva)$, where $\bva$ is the set of parameters. It is assumed that
	the prior densities of parameters $\bva$ and $\bpsi$ are $p(\bva)$ and $p(\bpsi)$, respectively.

	According to the principle of variational Bayesian inference,  we need specifying a family $\bmF$ of variational densities for random variables $\bbZ=\{\bbV,\bva,\bpsi\}$, which has the same support set as the posterior density $p(\bbZ|\bbD)$. To this end, let $q(\bbZ)\in\bmF$ be any variational density approximating the posterior density $p(\bbZ|\bbD)$. The main purpose of variational Bayesian inference is to find an optimal variational density $q^*(\bbZ)$ to approximate $p(\bbZ|\bbD)$ by minimizing the Kullback-Leibler divergence between $q(\bbZ)$ and $p(\bbZ|\bbD)$, which is equivalent to solving the following optimization problem $$q^*(\bbZ)=\arg\min_{q(\bbZ)\in\bmF}{\rm KL}\{q(\bbZ)||p(\bbZ|\bbD)\},$$
	where
	\begin{equation} \label{KL0}
		\begin{array}{llllll}
			{\rm KL}\{q(\bbZ)||p(\bbZ|\bbD)\} &=& \int\ln\left\{\dfrac{q(\bbZ)}{p(\bbZ|\bbD)}\right\}q(\bbZ){\rm d}\bbZ \\
			&=&\E_{q(\bbZ)}\{\ln q(\bbZ)\} - \E_{q(\bbZ)}\{\ln p(\by,\bbZ|\bbX)\} +\ln p(\by|\bbX),
		\end{array}
	\end{equation}
	in which $\E_{q(\bbZ)}(\cdot)$ represents the expectation taken with respect to the variational density $q(\bbZ)$. Note that ${\rm KL}\{q(\bbZ)||p(\bbZ|\bbD)\}$ is always larger than zero, and it is equal to zero if and only if $q(\bbZ)=p(\bbZ|\bbD)$. Let $L\{q(\bbZ)\}=\E_{q(\bbZ)}\{\ln p(\by,\bbZ|\bbX)\} - \E_{q(\bbZ)}\{\ln q(\bbZ)\}$. Thus, it follows from Equation (\ref{KL0}) that
	$$\ln p(\by|\bbX)={\rm KL}\{q(\bbZ)||p(\bbZ|\bbD)\}+L\{q(\bbZ)\} \geq L\{q(\bbZ)\},$$ which shows that $L\{q(\bbZ)\}$ can be seen as a lower bound for $\ln p(\by|\bbX)$, and is usually called the evidence lower bound (ELBO). Thus, minimizing ${\rm KL}(q(\bbZ)||p(\bbZ|\bbD))$ is equivalent to maximizing $L\{q(\bbZ)\}$ since $\ln p(\by|\bbX)$ does not depend on $\bbZ$. That is,
	\begin{equation}\label{QBZZ}
		q^*(\bbZ)=\arg\max_{q(\bbZ)\in\bmF} L\{q(\bbZ)\},
	\end{equation}
	which implies that finding the optimal variational density approximation problem of $p(\bbZ|\bbD)$ is transformed into maximizing $L\{q(\bbZ)\}$ over the variational family $\bmF$. It follows from Equation (\ref{KL0}) that
	the ELBO based on the joint probability density function $p(\by|\bbX,\bbV,\bpsi)$ and $p(\bbV|\bva)$
	has the form
	$$
	L\{q(\bbZ)\}=\int q(\bbZ)\ln\left\lbrace  \dfrac{p(\by|\bbX,\bbV,\bpsi)p(\bbV|\bva)p(\bva)p(\bpsi)}
	{q(\bbZ)}\right\rbrace {\rm d} \bbZ.
	$$
	
	It is rather difficult to directly implement the optimal problem (\ref{QBZZ}) due to the intractable high-dimensional integral involved. To address the issue, a widely used trick is to specify a simple variational density family $\bmF$, for example, the well-known mean-field variational density family. % are mutually independent and each is dominated by a distinct factor in the %variational density.
	That is, the variational density $q(\bbZ)$ is assumed to be factorized across the components of $\bbZ$:
	\begin{equation}\label{ELBO0}
		q(\bbZ)=q(\bbV)q(\bva)q(\bpsi)\equiv\prod_{s=1}^{S}q_s(\bthe_s),
	\end{equation}
	where $q_s(\bthe_s)$'s are variational densities of marginal components $\bthe_s$ in
	$\bbZ=\{\bthe_1,\ldots,\bthe_S\}$. Moreover, the optimal solutions of $q_s(\bthe_s)$'s are to be determined by maximizing $L\{q(\bbZ)\}$ with respect to the variational densities $q_1(\bthe_1),\ldots,q_S(\bthe_S)$ via the coordinate ascent method.

	Following
	the idea of the coordinate ascent method given in \cite{beal2003}, \cite{Bishop2007} and \cite{blei2018variational},
	when fixing %the variational densities $q_j(\bthe_j)$ of
	other factors $\bthe_j$ for $j\neq s$, i.e., $\bthe_{-s}=\{\bthe_j: j=1,\ldots,s-1,s+1,\ldots,S\}$, it is easily shown from Equations (\ref{QBZZ}) and (\ref{ELBO0}) that the optimal density $q^*_s(\bthe_s)$ %that maximizes $L\{q(\bbZ)\}$ with respect to $q_s(\bthe_s)$ %is shown to take the form
	has the form
	\begin{equation}\label{calcf1}
		q^*_s(\bthe_s)\propto \exp[\E_{-s}^*\{\ln p(\bfm{y}|\bbX,\bbV,\bpsi) + \ln p(\bbV|\bva)+ \ln p(\bva)+\ln p(\bpsi)\}],
	\end{equation}
	where $\E_{-s}^*(\cdot)$ represents the expectation taken with respect to the optimal variational density $q^*_{-s}(\bthe_{-s})$
	of $\bthe_{-s}$.
	Equation (\ref{calcf1}) indicates that the expectation $\E_{-s}^*(\cdot)$ does not depend on the optimal variational density of $\bthe_{s}$, and the optimal variational density $q^*_s(\bthe_s)$ cannot be obtained once in that the variational densities $q_{-s}^*(\bthe_{-s})$ are not the optimal ones. To address the issue, the widely used coordinate updating algorithm is employed to iteratively update $q_s^*(\bthe_s)$ via Equation (\ref{calcf1}). Once the coordinate updating algorithm converges, one can regard
	$q_s^*(\bthe_s)$ as the desirable optimal variational density and take its posterior mode as variational Bayesian estimate of $\bthe_s$.

	\section{Variational Bayesian Estimation on Logistic Tensor Regression}  %%% Section 3
	
	\subsection{Logistic Tensor Regression} \label{subsect3.1}
	
	Consider a dataset $\bbD=\{(\bmX_i,y_i): i=1,2,\cdots,n\}$ with $n$ subjects, where
	$y_{i}\in\{-1,1\}$ is an observation of binary response variable $Y$ for the $i$th subject
	indicating that some event occurs or not (e.g., the $i$th subject has some disease),
	and $\bmX_i\in\bbR^{I_{1} \times I_{2} \times \cdots \times I_{M}}$ is a $M$th-order tensor ($M\ge 2$)
	associated with some image, which can be regarded as covariates having the effect on response variable.
	Here we utilize the following logistic tensor regression
	\begin{equation}\label{equ1}
		\Pr\left(Y=y_{i} \mid \bmX_{i}, \bmW\right)
		=
		\dfrac{1}{1+\exp\left\{-y_{i}\left\langle\bmW, \bmX_{i}\right\rangle\right\}}
		\equiv g\left(y_{i}\left\langle\bmW, \bmX_{i}\right\rangle\right),
	\end{equation}
	to fit the considered dataset $\bbD$,
	where $\bmW\in\bbR^{I_{1}\times I_{2}\times\cdots\times I_{M}}$ is a tensor of regression coefficients,
	indicating that the number of the unknown parameters in the considered model is $\prod_{j=1}^{M}I_j$, and
	$g(a)=\{1+\exp(-a)\}^{-1}$ is called logistic function.
	%If the probability $\Pr(y_i=1|\bmX_i,\bmW)$ is larger than a given threshold value for the decision, the label of the $i$th image is %detected to be $1$, otherwise $-1$.
	To estimate unknown parameters $\bmW$ based on the observed dataset $\bbD$, one may consider
	vectorization of tensor $\bmW$, which can not retain the intrinsic spatial correlation structure of tensor data
	and often leads to the biased estimators of parameters and the poor performance of prediction.
	%We require considering the whole structure of the tensor.
	Moreover, due to the high-dimensionality of tensor $\bmW$, %the unknown parameters in tensor coefficient are high-dimensional obviously.
	it is quite troublesome to directly estimate $\bmW$, which may lead to the overfitting problem.
	To solve the aforementioned problem, we first consider the low-rank decomposition approximations to $\bmW$, and then utilize the preceding introduced variational Bayesian method to evaluate variational Bayesian estimates of parameters in the approximated model of logistic tensor regression.

	First, it follows from Equation (\ref{unfolded}) that model (\ref{equ1}) can be expressed equivalently as
	$$%\begin{equation}\label{logistic_matrix}
	g\left(y_{i}\left\langle\bmW, \bmX_{i}\right\rangle\right)=g\left(y_{i}\textmd{trace}\{\bW_{(1)}\bX_{i(1)}^{\top}\}\right)
	=\cdots=g\left(y_{i}\textmd{trace}\{\bW_{(M)}\bX_{i(M)}^{\top}\}\right),
	$$%\end{equation}
	where $\bW_{(m)}$ and $\bX_{i(m)}$ are the mode-$m$ matricization versions of tensors $\bmW$ and $\bmX_i$ for $m\in\{1,2,\ldots,M\}$, respectively.
	By the CP decomposition of tensor \citep{carroll70,Hars1970}, coefficient tensor $\bmW$ can be approximated by $\bmW\approx\sum_{r=1}^{R}\bu_{r}^{(1)}\circ\bu_{r}^{(2)}\circ\cdots\circ\bu_{r}^{(M)}\equiv \bfm{\mathcal{C}}$ under the low-rank assumption, and
	the mode-$m$ matricization of the low-rank tensor $\bfm{\mathcal{C}}$ has the form
	$$\bfm{C}_{(m)}=\bU^{(m)}\left(\bU^{(M)} \odot \cdots \odot \bU^{(m+1)} \odot \bU^{(m-1)} \odot \cdots \odot \bU^{(1)}\right)^{\top}.$$
	Thus, the mode-$m$ matricization of $\bmW$ can be approximated by
	\begin{equation}\label{equw2}
		\bW_{(m)}\approx\bfm{C}_{(m)}=\bU^{(m)}\bU^{(-m)^{\top}},
	\end{equation}
	where $\bU^{(m)}=\left(\bu_{1}^{(m)},\bu_{2}^{(m)},\ldots,\bu_{R}^{(m)}\right)\in\bbR^{I_{m}\times R}$, and $\bU^{(-m)}=\bU^{(M)}\odot\bU^{(M-1)}\odot\cdots\odot\bU^{(m+1)}\odot\bU^{(m-1)}\odot\cdots\odot\bU^{(1)}$ for $m=1,2,\ldots,M$.		
	 Equation (\ref{equw2}) shows that the mode-$m$ matricization of $\bmW$ only depends on its CP decomposition rather than the selected mode $m$. That is, the mode-$m_1$ matricization of $\bmW$ has the same components as its mode-$m_2$ matricization for any $m_1\neq m_2\in\{1,\ldots,M\}$ in that matricization of tensor is just the process of reordering components of tensor into a matrix but does not change values of components in tensor.
		Thus, choosing different modes for matricization leads to the same results \citep{carroll70,Hars1970}.
%	By Equation (\ref{equw2}), 
	Thus, we obtain
	$$\textmd{trace}\{\bW_{(m)}\bX_{i(m)}^{\top}\}\approx\textmd{trace}\{\bU^{(m)}\bU^{(-m)^{\top}}\bX_{i(m)}^{\top}\}
	=\textmd{vec}(\bU^{(m)})^{\top}(\bI_R\otimes\bX_{i(m)})\textmd{vec}(\bU^{(-m)}),$$
	where the symbol $\otimes$ represents the Kronecker product of two matrices.	
	Thus, logistic tensor regression (\ref{equ1}) for any $m\in\{1,2,\cdots,M\}$ can be approximated by
	\begin{equation}\label{equ3}
		\Pr\left(Y=y_{i} \mid \bmX_{i}, \bmW\right)
		%g\left(y_{i}\textmd{trace}\{\bW_{(m)}\bX_{i(m)}^{\top}\}\right)
		\approx
		g\left\{y_{i}\textmd{vec}(\bU^{(m)})^{\top}(\bI_R\otimes\bX_{i(m)})\textmd{vec}(\bU^{(-m)})\right\},
	\end{equation}		
	which implies that we only requires estimating $R\sum_{j=1}^{M}I_j$ parameters.
	 That is, the CP decomposition of $\bmW$ can largely reduce
		the number of parameters to be estimated in the approximated model (\ref{equ3}) from $\prod_{j=1}^{M}I_j$ to $R\sum_{j=1}^{M}I_j$ with the appropriately selected value of $R$, which is much less than the dimensionality
		of the model (\ref{equ1}). Thus, the selection of $R$ is of critical importance for dimension reduction of CP decomposition.
		The details for selecting the optimal $R$ value refer to Section 3.4 below. Since $\bU^{(m)}$ and $\bU^{(-m)}$ are separative under the mode-$m$ matricization of $\bmW$ for any $m\in\{1,\ldots,M\}$, we can construct some estimator of low dimensional factor matrix $\bU^{(m)}$ (i.e., its corresponding dimension $RI_m$ is much less than the dimension $\prod_{j=1}^{M}I_j$ of $\bmW$) given factor matrix $\bU^{(-m)}$, which shows that the approximation procedure introduced above can largely facilitate the estimation of $\bmW$.

	Let $\bbU=\{\bU^{(j)}: j=1,2,\ldots,M\}$ be a set of unknown parameters.
	Denote $\by=(y_1,y_2,\ldots,y_n)^{\top}$ as a vector of $n$ observations for response variable $Y$, and
	$\bbX=\{\bmX_i: i=1,2,\ldots,n\}$ as a set of $n$ image data.
	%Let $\bthe^{(j)}=\textmd{vec}(\bU^{(j)})\in\bbR^{I_{j}R}$,
	%and $\bthe=(\bthe^{(1)\top}, \bthe^{(2)\top}, \cdots, \bthe^{(M)\top})^{\top}\in\bbR^{R\sum_{j=1}^{M}I_j}$.
	Under the above notation, the joint probability mass function of $\by$ in the model (\ref{equ3}) is given by
	\begin{equation}\label{UAPP}
		\begin{array}{lll}
			& p\left(\by \mid \bbX, \bmW\right)=\prod\limits_{i=1}^n\Pr(Y=y_i|\bmX_i,\bmW)\\
			&\approx \prod\limits_{i=1}^n g\left(y_{i}\textmd{vec}(\bU^{(m)})^{\top}(\bI_R\otimes\bX_{i(m)})\textmd{vec}(\bU^{(-m)})\right)
			\equiv p\left(\by \mid \bbX, \bbU,m\right),
		\end{array}
	\end{equation}
	%			Equation (\ref{UAPP})
	which shows that we can easily estimate the parameter matrix $\bU^{(m)}$ when $\bU^{(-m)}$ is given in that $\bU^{(m)}$ and $\bU^{(-m)}$ are separative for any $m\in\{1,2,\ldots,M\}$.
	% we can easily estimate the parameter matrices in the set $\bbU$. As we estimate the matrix $\bU^{(j)}$, we consider the probability $\Pr\left(\by %\mid \bbX, \bbU,m=j\right)$, where $\bU^{(-m)}$ does not include $\bU^{(m)}$ and $\bU^{(m)}$ has been separated from $\bU^{(-m)}$, which can avoid %complex derivation and operation from the $\bU^{(-m)}$.
	
	Although we reduce the dimensionality of the model (\ref{equ1}) from $\prod_{j=1}^{M}I_j$ to $R\sum_{j=1}^{M}I_j$ based on the approximation procedure introduced above, there are still two challenges in estimating parameters $\bbU$:
	(i) $R\sum_{j=1}^{M}I_j$ might be still larger than
	the sample size in that image data might have a relatively large pixel size; % or tensor data might have high order and high dimension of each mode.
	(ii) parameter matrix $\bU^{(m)}$ might be sparse for $m=1,2,\ldots,M$.
	%Moreover, we are interested in identifying geometric sub-regions of the tensor across where coefficients are not close to zero, with %the remaining elements being very close to zero. 	
	To solve these challenges, one can estimate parameters $\bbU$ via the regularization method with the Lasso penalty \citep{Tan2013} or adaptive Lasso penalty \citep{Zhou2013}.
	%	which could be interpreted as posterior mode estimates based on Bayesian lasso method \citep{Park2008}.
	However, the regularization method heavily depends on the selection of the tuning parameter, which might be sensitive to
	the dimensionality of tensor, the degree of sparsity and the CP rank of tensor.
	As an alternative to the regularization method, $\bbU$ can be estimated by
	%the traditional Bayesian approaches such as Markov chain Monte Carlo, or
	the penalized Bayesian approaches such as Bayesian Lasso and Bayesian adaptive Lasso \citep{pmlr-v37-suzuki15,JMLR:v18:16-362,Yi2022Tang} via Markov chain Monte Carlo (MCMC) algorithm
	by introducing shrinkage priors \citep{pmlr-v37-suzuki15,JMLR:v18:16-362}, which can effectively shrink
	unimportant parameters to zero in estimating unknown parameters $\bbU$.

	\subsection{Prior Distributions}
	
	To make Bayesian inference on $\bbU$ based on the probability mass function (\ref{UAPP}),
	we require specifying the priors of unknown parameters in the set $\bbU$.
	In high-dimensional and sparse issue, traditional method for specifying the priors of unknown parameters
	might lead to the ill-posed problem, i.e., not able to induce shrinkage within and across components
		in tensor factorization of tensor coefficient for optimal high-dimensional region selection
		and  favor recovery of contiguous geometric subregions of tensor coefficient \citep{JMLR:v18:16-362}.
	To address the issue, we employ the multiway Dirichlet generalized double Pareto (M-DGDP) prior \citep{JMLR:v18:16-362} to induce the shrinkage components of $\bbU$.
	The M-DGDP prior can accommodate the shrinkage of the tensor coefficient for the appropriate identification of important components in the tensor predictors and enable high dimensional region selection.
	Following \citep{JMLR:v18:16-362}, the hierarchical margin-level M-DGDP priors for $\bu^{(j)}_r$ ($r=1,\ldots,R; j=1,\ldots,M$) can be expressed as
	\begin{equation}\label{proir1}
		\bu^{(j)}_r\sim N(\0,\tau\phi_r\bSi_{jr}), \tau \sim \Gamma(a_{\tau}, b_{\tau}),
	\end{equation}
	\begin{equation}\label{proir2}
		\bSi_{jr}=\textmd{diag}(\sigma_{jr1},\sigma_{jr2},\cdots,\sigma_{jrI_j}),
		\sigma_{jrk}\sim {\rm Exp}(\lambda_{jr}^2/2), \lambda_{jr}\sim \Gamma(a_{\lambda},b_{\lambda}),
	\end{equation}
	\begin{equation}\label{proir3}
		\bPhi=(\phi_1,\phi_2,\cdots,\phi_R)^{\T}\sim {\rm Dirichlet}(\alpha_1,\alpha_2,\cdots,\alpha_R),\;\;
		\sum\nolimits_{r=1}^R\phi_r=1,
	\end{equation}	
	where $\tau\phi_r$ is the global scale for each component, $\bPhi$ is used to shrinkage towards lower ranks in the assumed CP decomposition, and $\bSi_{jr}$ is the margin-specific scale for each component, $\Gamma(\cdot,\cdot)$ is the gamma distribution,
	Dirichlet($\cdot,\cdots,\cdot$) denotes the Dirichlet distribution, and Exp($\cdot$) represents the exponential distribution with parameter $a$. Under the above specified priors, it is easily shown that
	$u^{(j)}_{rk}|\lambda_{j r},\phi_{r},\tau\sim {\rm DE}(\lambda_{j r}/\sqrt{\phi_{r}\tau})$ for $k=1,\ldots,I_j$,
	where DE($\cdot$) denotes the double-exponential distribution, $u_{rk}^{(j)}$ is the $k$th component of $\bu_r^{(j)}$.
	The above specified prior yields the generalized double Pareto prior on the margin coefficients $\bu^{(j)}_r$, which consequently
	has the form of the adaptive Lasso penalty function.
	Flexibility in estimating $\bu^{(j)}_r$ is accommodated by modeling within-margin heterogeneity via
	element-specific scaling $\sigma_{jrk}$. The common rate parameter $\lambda_{jr}$ shares the information between
	margin components, which is beneficial for shrinkage at the local scale. The M-DGDP prior can impart shrinkage of the tensor components at global and local levels, while also encouraging shrinkage towards low rank tensor decomposition \citep{JMLR:v18:16-362}.

	Denote $\bsi_{(j)}=(\sigma_{j11}, \sigma_{j12},\cdots,\sigma_{j1I_{j}},\cdots,\sigma_{jR1}, \sigma_{jR2},\cdots,\sigma_{jRI_{j}})^{\top}$,  $\bsi=(\bsi_{(1)}^{\top},\cdots,\\ \bsi_{(M)}^{\top})^{\top}$, and $\bla=(\lambda_{11},\lambda_{12},\cdots,\lambda_{MR})^{\top}$.
	Denote $\bva=\left\{\bsi, \bPhi,  \bla, \tau\right\}$ as the set of parameters associated with the priors (\ref{proir1})--(\ref{proir3}). The hyper-parameters $a_{\tau}, b_{\tau}, a_{\lambda}, b_{\lambda}$ and $\bal=(\alpha_1,\alpha_2,\ldots,\alpha_R)^{\top}$ are the pre-specified or randomly selected in some pre-specified interval \citep{Chow2011ET}.
	Generally, these hyper-parameters can be fixed at some appropriate values or estimated via empirical Bayesian approach.
	Similarly to \citep{JMLR:v18:16-362}, we set these hyper-parameters as $\alpha_1=\cdots=\alpha_R\equiv\alpha=1/R$, $a_{\tau}=\alpha R$,
	$b_{\tau}=0.5R\sum_{j=1}^MI_j-0.5$, $a_{\lambda}=3$, and $b_{\lambda}=a_{\lambda}^{0.5/M}$ in our simulation studies and real example analysis. % where the choices for hyper-parameters $a_{\lambda}$ and $b_{\lambda}$ are not sensitivity.

	Note that the CP decomposition may suffer from the identification problem for the margins $\bu^{(j)}_r$ in Equation (\ref{equ3}), which may come from three sources: the scale identification, permutation identification and orthogonal transformation identification \citep{JMLR:v18:16-362,Billio2023}. Although the margins $\bu^{(j)}_r$ are unidentifiable, the tensor coefficient $\bmW$ is identifiable. Under the M-DGDP priors for tensor margins, we can make Bayesian inference on $\bmW$ \citep{JMLR:v18:16-362} in terms of their joint posterior distribution. Thus we can avoid imposing identifiability restrictions on tensor margins.
	%As is evident from our numerical studies, non-identifiability of tensor margins does not appear to affect convergence and performance.
	
	 With the shrinkage prior introduced above, the Bayesian approaches via MCMC algorithm can be developed.
		However, the MCMC-based approaches suffer from several drawbacks mentioned in Introduction, and have low convergence rate for analyzing tensor data related with images. To address these issues, in what follows, we develop a variational Bayesian method to simultaneously identify the important tensor predictors and estimate unknown parameters $\bbU$ with some appropriate shrinkage priors.
		The proposed variational Bayesian method avoids the sampling problem involved in the MCMC algorithm, is orders of magnitude faster than the MCMC algorithm for achieving the same accuracy of parameter estimation, and requires the smaller memory than the MCMC algorithm.  	
		Also, numerical studies conducted in Section 4 show that the proposed variational Bayesian method is
		computational inexpensive and more efficient than the MCMC algorithm.
	
	%\subsection{Variational Bayesian Estimation}
	\subsection{Variational Bayesian Estimation of $\bbU$}\label{section33}
	
	It follows from Equations (\ref{UAPP}) and (\ref{proir1}) that the posterior density function of $\bbU$ given the data ($\by,\bbX$) and parameters
	$\{\tau,\bPhi,\bsi\}$ can be expressed as
	$$
	\begin{array}{lll}
		&&p(\bbU|\by,\bbX,\tau,\bPhi,\bsi,m)\propto p(\by|\bbX,\bbU,m)p(\bbU|\tau,\bPhi,\bsi)\\
		&&\approx \left\{\prod\limits_{i=1}^ng\left(y_{i}\textmd{vec}(\bU^{(m)})^{\top}(\bI_R\otimes\bX_{i(m)})\textmd{vec}(\bU^{(-m)})\right)\right\}
		\left\{\prod\limits_{r=1}^R\prod\limits_{j=1}^Mp(\bu_r^{(j)}|\tau,\phi_r,\bSi_{jr})\right\},
	\end{array}
	$$
	which indicates that it is rather difficult to make variational Bayesian inference on $\bbU$ with $p(\bbU|\by,\bbX,\tau,\bPhi,\bsi,m)$
	due to nonlinear function $g(\cdot)$ involved. To overcome the difficulty, in what follows, we consider the approximation to $g(\cdot)$ or $p(\by|\bbX,\bbU,m)$.
		Since $g(a)=\{1+\exp(-a)\}^{-1}$, we have
		\begin{equation}\label{loglogistic}
			\log\{g(a)\}=-\log\{1+\exp(-a)\}=\dfrac{a}{2}-\log\{\exp(a/2)+\exp(-a/2)\}\equiv \dfrac{a}{2}+f(a),
		\end{equation}
		where $f(a)=-\log\{\exp(a/2)+\exp(-a/2)\}$ is a convex non-linear function in the variable $a^2$.
		Following the argument of \cite{Jaakkola2000BayesianPE}, we can bound $f(a)$ globally with a first-order Taylor expansion of $f(a)$
		in the variable $a^2$, i.e.,
		\begin{equation}\label{inequation_f}
			f(a)\geqslant f(\xi)+\dfrac{d f(\xi)}{d\xi^2} (a^2-\xi^2)=-\dfrac{\xi}{2}+\log g(\xi)-\lambda(\xi)(a^2-\xi^2),
		\end{equation}
		which is a linear function in the variable $a^2$,
		where $\lambda(\xi)=(g(\xi)-1/2)/(2 \xi)$, and $\xi$ is a variational parameter corresponding to $a$.
		Combining (\ref{loglogistic}) and (\ref{inequation_f}) yields
		$$			\log\{g(a)\}\geqslant \log\{ g(\xi)\} +\frac{a-\xi}{2} - \lambda(\xi)(a^2 - \xi^2),	$$
		which leads to
		\begin{equation}\label{gzw}
			g(a) \geqslant g(\xi) \exp \left\{(a-\xi)/2-\lambda(\xi)(a^{2}-\xi^2)\right\}\equiv \ell(a;\xi).
		\end{equation}
		The lower bound $\ell(a;\xi)$ is no longer normalized, and the approximation is a tight lower bound on the sigmoid with an additional parameter
		$\xi$ \citep{Jaakkola2000BayesianPE}.
		The lower bound, which has the Gaussian or exponential form of a quadratic function in the variable $a$, is an alternative to the logistic function, i.e., a variational transformation of the logistic function. Thus, based on the priors and the variational transformation likelihood, we can obtain a Gaussian approximation to the posterior. Note that the lower bound of the variational transformation is equivalent to the Pl$\acute{\rm o}$ya-Gamma data augmentation approach \citep{Polson2013,Daniele2019}, and the variational transformation method can avoid estimating the optimal density of the extra parameters in the Pl$\acute{\rm o}$ya-Gamma data augmentation approach.
		Their details are presented in the Supplementary Material \citep{Jin2025}.

	Based on Equation (\ref{gzw}), we can obtain  
	the lower bound of $\ln p(\by| \bbX, \bbU, m)$:
	\begin{equation}\label{APPGZ}
		\begin{array}{llllll}
			& &\ln p(\by| \bbX, \bbU, m) \\
			&\geqslant&
%			\sum\limits_{i=1}^{n}\left\{\ln g(\xi_i)-\dfrac{\xi_i}{2}+\lambda(\xi_i) \xi_i^{2}\right\}  +\dfrac{1}{2}\textmd{vec}(\bU^{(m)})^{\top}\sum\limits_{i=1}^{n}y_i(\bI_R\otimes\bX_{i(m)})\textmd{vec}(\bU^{(-m)}) \\ 
%			&-&\textmd{vec}(\bU^{(m)})^{\top} \sum\limits_{i=1}^{n}\lambda(\xi_i)(\bI_R\otimes\bX_{i(m)})\textmd{vec}(\bU^{(-m)})\textmd{vec}(\bU^{(-m)})^{\top}
%			(\bI_R\otimes\bX_{i(m)}^{\top})\textmd{vec}(\bU^{(m)}) \\ 
%			&=&
% 			\ln h(\bbU, \bxi|m)&=&
 \sum\limits_{i=1}^{n}\left\{\ln g(\xi_i)-\dfrac{\xi_i}{2}+\lambda(\xi_i) \xi_i^{2}\right\}
 +\textmd{vec}(\bU^{(m)})^{\top}\ba^{(-m)}
 -\textmd{vec}(\bU^{(m)})^{\top}\bOm^{(-m)}\textmd{vec}(\bU^{(m)})\\
			&\equiv& \ln h(\bbU, \bxi|m),
		\end{array}
	\end{equation}
where
	 $\ba^{(-m)}=\{\sum_{i=1}^{n}y_i(\bI_R\otimes\bX_{i(m)})\}\textmd{vec}(\bU^{(-m)})/2$, $\bxi=(\xi_1,\cdots,\xi_n)^{\T}$ is a vector of local variation parameters,  and $$\bOm^{(-m)}=\sum_{i=1}^{n}\lambda(\xi_i)(\bI_R\otimes\bX_{i(m)})\textmd{vec}(\bU^{(-m)})\textmd{vec}(\bU^{(-m)})^{\top}(\bI_R\otimes\bX_{i(m)}^{\top}).$$
	Combining the lower bound of ${\rm ln} p(\by|\bbX,\bbU,m)$ and the definition of the ELBO given in
	(\ref{QBZZ}), we can build the following ELBO:
	\begin{equation}\label{equw4}
		L\{q(\bbZ), \bxi|m\}=\int q(\bbZ) \ln \left\lbrace \dfrac{h(\bbU,\bxi|m) p(\bbU | \bva)p(\bva)}
		{q(\bbZ)} \right\rbrace {\rm d} \bbZ,
	\end{equation}
	where
	%Based on mean-field variational theory,
	the variational density 
	$q(\bbZ)$ is assumed to be factorized across the components of $\bbZ$:
	\begin{equation}\label{MFV01}
		q(\bbZ)=q(\tau)q(\bPhi)\prod_{j=1}^M\prod_{r=1}^Rq(\bu_r^{(j)})\prod_{j=1}^{M}\prod_{r=1}^R\prod_{k=1}^{I_j}q(\sigma_{jrk})\prod_{j=1}^M\prod_{r=1}^Rq(\lambda_{jr})
		\equiv\prod_{s=1}^{S}q_s(\bthe_s).
	\end{equation}
	%By maximizing $L_{\ell}\{q(\bbZ),\bxi\}$ with respect to $q_s(\bthe_s)$,
	%			Combining Equations (\ref{calcf1}) and (\ref{equw4}),
	In the framework of Equations (\ref{calcf1}),
	we can obtain
	the following optimal variational density of $\bthe_s$:
	\begin{equation}\label{calcf01}
		q^*_s(\bthe_s)\propto \exp[\E_{-s}^*\{\ln h(\bbU,\bxi|m) + \ln p(\bbU|\bva)+ \ln p(\bva)\}].
	\end{equation}

	It follows from Equations (\ref{APPGZ}) and (\ref{calcf01}) together with the prior (\ref{proir1}) of $\bu_r^{(j)}$ that
	the optimal variational density $q^*(\bu_r^{(j)})$ for $\bu_r^{(j)}$ has the form
	\begin{equation}\label{qu}
		q^*(\bu_r^{(j)})\sim N\left(\bmu_{\bu_r^{(j)}},\bSi_{\bu_r^{(j)}}\right),
	\end{equation}
	where
	$$
	\begin{array}{rllll}	\bSi_{\bu_r^{(j)}}^{-1}&=&2\E^*_{\bU^{(-j)}}\{\bOm^{(-j)}\}_{rr}+\E^*_{\tau}(\tau^{-1})\E^*_{\phi_r}(\phi_r^{-1})
		\E^*_{\bsi}(\bSi_{jr}^{-1}),\\
		\bmu_{\bu_r^{(j)}}&=&\bSi_{\bu_r^{(j)}}\left[ \E^*_{\bU^{(-j)}}\{\ba^{(-j)}\}_r-2\sum_{\ell\neq r}\E^*_{\bU^{(-j)}}\{\bOm^{(-j)}\}_{r\ell}\E^*_{\bu_{\ell}^{(j)}}\{\bu_{\ell}^{(j)}\}\right],\\
		\E^*_{\bU^{(-j)}}\{\ba^{(-j)}\}&=&\dfrac{1}{2}\sum\limits_{i=1}^{n}y_i(\bI_R\otimes\bX_{i(j)})\textmd{vec}(\E^*_{\bU^{(-j)}}\{\bU^{(-j)}\}),\\
		\E^*_{\bU^{(-j)}}\{\bOm^{(-j)}\}&=&\sum\limits_{i=1}^{n}\lambda(\xi_i)(\bI_R\otimes\bX_{i(j)})\E^*_{\bU^{(-j)}}\left\lbrace \textmd{vec}(\bU^{(-j)})\textmd{vec}(\bU^{(-j)})^{\top}\right\rbrace (\bI_R\otimes\bX_{i(j)}^{\top}).
	\end{array}
	$$
	in which
	$\E^*_{\bU^{(-j)}}\{\bOm^{(-j)}\}_{r\ell}$ is a $I_j\times I_j$ matrix from the $(r,\ell)$th block matrix of the
	matrix $\E^*_{\bU^{(-j)}}\{\bOm^{(-j)}\}$,
	$\E^*_{\bU^{(-j)}}\{\ba^{(-j)}\}_r$ is a $I_j\times 1$ vector from the $r$th subvector of the vector
	$\E^*_{\bU^{(-j)}}\{\ba^{(-j)}\}$ corresponding to $\bu_r^{(j)}$.
	Thus, the approximate posterior mean and second moment of $\bu_r^{(j)}$ are given by
	$\E^*_{\bu}(\bu_r^{(j)})=\bmu_{\bu_r^{(j)}}$ and $\E^*_{\bu}(\bu_r^{(j)}\bu_r^{(j)\top})=\bSi_{\bu_r^{(j)}}+\bmu_{\bu_r^{(j)}}\bmu_{\bu_r^{(j)}}^{\top}$, for $j=1,\ldots,M$ and $r=1,\ldots,R$.
	
	For the computation of $\E^*_{\bU^{(-j)}}\{\bU^{(-j)}\}$ and $\E^*_{\bU^{(-j)}}\{\textmd{vec}(\bU^{(-j)})\textmd{vec}(\bU^{(-j)})^{\top}\}$,
	it is easily shown from the mean-field theory and matrix operation that
	$$
	\begin{array}{lllllll}
		& &\E^*_{\bU^{(-j)}}\{\bU^{(-j)}\}=\\
		& &
		\E^*_{\bU^{(-j)}}\{ \bU^{(M)}\}\odot\cdots\odot\E^*_{\bU^{(-j)}}\{\bU^{(j+1)}\}\odot\E^*_{\bU^{(-j)}}\{\bU^{(j-1)}\}\odot\cdots
		\odot\E^*_{\bU^{(-j)}}\{\bU^{(1)}\},
	\end{array}
	$$
	where $\E^*_{\bU^{(-j)}}\{\bU^{(-j)}\}$ is a $(\prod_{i=1}^{j-1}I_i\prod_{i=j+1}^M I_i)\times R$ matrix,
	and its $r$th column vector $(r=1,2,\ldots,R)$ can be calculated by
	$$\E_{\bu_r^{(M)}}^*(\bu_r^{(M)})\otimes\cdots\otimes\E_{\bu_r^{(j+1)}}^*(\bu_r^{(j+1)})\otimes
	\E_{\bu_r^{(j-1)}}^*(\bu_r^{(j-1)})\cdots\otimes\E_{\bu_r^{(1)}}^*(\bu_r^{(1)}).$$
	Similarly,
	$\E^*_{\bU^{(-j)}}\{\textmd{vec}(\bU^{(-j)})\textmd{vec}(\bU^{(-j)})^{\top}\}$ is a $(R\prod\limits_{i=1}^{j-1}I_i\prod\limits_{i=j+1}^M I_i)\times(R\prod\limits_{i=1}^{j-1}I_i\prod\limits_{i=j+1}^M I_i)$ matrix, and can be written as an $R\times R$ block matrix
	whose $(k,t)$th block matrix is a $(\prod_{i=1}^{j-1}I_i\prod_{i=j+1}^M I_i)\times(\prod_{i=1}^{j-1}I_i\prod_{i=j+1}^M I_i)$ matrix
	and has the form:
	$$
	\begin{array}{llllll}
		\E_{\bu_{t}^{(M)}}^*(\bu_{t}^{(M)}\bu_{t}^{(M)\T})
		\otimes\cdots
		\otimes
		\E_{\bu_{t}^{(j+1)}}^*(\bu_{t}^{(j+1)}\bu_{t}^{(j+1)\T})\\
		\otimes
		\E_{\bu_{t}^{(j-1)}}^*(\bu_{t}^{(j-1)}\bu_{1}^{(j-1)\T})\otimes
		\cdots\otimes
		\E_{\bu_{t}^{(1)}}^*(\bu_{t}^{(1)}\bu_{t}^{(1)\T})
	\end{array}
	$$
	for $k=t$, and
	$$
	\begin{array}{lll}
		&\E_{\bu_{k}^{(M)}}^*(\bu_{k}^{(M)})
		\E_{\bu_{t}^{(M)}}^*(\bu_{t}^{(M)\T})\otimes\cdots\otimes
		\E_{\bu_{k}^{(j+1)}}^*(\bu_{k}^{(j+1)})
		\E_{\bu_{t}^{(j+1)}}^*(\bu_{t}^{(j+1)\T})\\ 
		&\otimes
		\E_{\bu_{k}^{(j-1)}}^*(\bu_{k}^{(j-1)})$
		$\E_{\bu_{t}^{(j-1)}}^*(\bu_{1}^{(j-1)\T})\otimes\cdots\otimes
		\E_{\bu_{k}^{(1)}}^*(\bu_{k}^{(1)})
		\E_{\bu_{t}^{(1)}}^*(\bu_{t}^{(1)\T})
	\end{array}
	$$ for $k\not=t$.
	
	Now consider the optimal variational density for $\tau$.
	By Equations (\ref{APPGZ}) and (\ref{calcf01}) and the prior of $\tau$ given in (\ref{proir1}),
	it is easily shown  that
	the optimal variational density $q^*(\tau)$ for $\tau$ is the generalized inverse Gaussian distribution:
	\begin{equation}\label{qtau}
		q^*(\tau) \sim {giG} \left( P_{\tau}, \alpha_{\tau}, \beta_{\tau}\right),
	\end{equation}
	where 
	$\beta_{\tau}=\sum_{j=1}^{M}\sum_{r=1}^R
	{\rm trace}\left\lbrace \E^*_{\phi_r}(\phi_r^{-1})\E^*_{\bsi}(\bSi_{jr}^{-1})
	\E^*_{\bu_r^{(j)}}(\bu_r^{(j)}\bu_r^{(j)\top})
	\right\rbrace$, $\alpha_{\tau}=2a_{\tau}$ and $P_{\tau}=b_{\tau}-R\sum_{j=1}^{M}I_{j}/2$.

	Now consider the optimal variational density $q^*(\sigma_{jrk})$ for $\sigma_{jrk}$.
	Again, it follows from Equations (\ref{APPGZ}) and (\ref{calcf01}) and the prior of $\sigma_{jrk}$ given in (\ref{proir2}) that
	$q^*(\sigma_{jrk})$ is also the generalized inverse Gaussian distribution:
	\begin{equation}\label{qsigma}
		q^*(\sigma_{jrk})\sim giG(P,a_{jr},b_{jrk}),
	\end{equation}
	where $P=1/2$, $a_{jr}=\E^*_{\lambda_{jr}}(\lambda_{jr}^{2})$,
	and
	$b_{jrk}=\E^*_{\tau}(\tau^{-1})\E^*_{\phi_r}(\phi_r^{-1})\E^*_{u_{rk}^{(j)}}\{(u_{rk}^{(j)})^2\}.$
	
	%			Again, $\E^*_{\sigma_{jrk}}\{\ln(\sigma_{jrk})\}$ can be approximated by $\ln\{\E^*_{\sigma_{jrk}}(\sigma_{jrk})\}$.

	For the optimal variational density $q^*(\phi_r)$ for $\phi_r$ ($r=1,\ldots,R$), without the constraint
		$\sum_{r=1}^R\phi_r=1$,
	it follows from Equations (\ref{APPGZ}) and (\ref{calcf01}) and the prior of $\phi_r$ given in (\ref{proir3}) that
	$q^*(\phi_r)$ is the inverse gamma distribution:
	\begin{equation}\label{qphi}
		q^*(\phi_{r}) \sim  iGamma\left(a_{r}, b_{r}\right),
	\end{equation}
	where
	$b_{r}=\frac{1}{2}\E^*_{\tau}(\tau^{-1})\sum_{j=1}^{M}{\rm trace}
	\left\lbrace \E^*_{\bsi}(\bSi_{jr}^{-1})\E^*_{\bu_r^{(j)}}(\bu_{r}^{(j)}\bu_{r}^{(j)\top})\right\rbrace$ and	$a_{r}=\frac{1}{2} \sum_{j=1}^{M}I_{j}-\alpha_{r}$.
		Under the constraint $\sum_{r=1}^{R}\phi_r=1$, we let $\tilde\phi_r=b_r\phi_r^{-1}/(\sum_{j=1}^Rb_j\phi_j^{-1})$.
		It is easily shown that the joint variational density of $\tilde\phi_1,\ldots,\tilde\phi_R$ still follows the Dirichlet distribution, i.e.,
		Dirichlet($a_1,\ldots,a_R$), and the variational density of $\tilde\phi_r$ follows the Beta distribution with parameters $a_r$ and $\sum_{j\neq r}a_j$. Thus, we have the approximate posterior moments of $\phi_r$: $\E^*_{\phi_r}\{\ln(\phi_r)\}=\ln(a_r)-\varpi(\sum_{j=1}^R a_j)$, where $\varpi(\cdot)$ is the digamma function, and the sample moments $\E^*_{\phi_r}(\phi_r^{-1})$ and $\E^*_{\phi_r}(\phi_r^{1/2})$ based on the updating variational densities of $\phi_r$.

	Since $u^{(j)}_{rk}|\lambda_{j r},\phi_{r},\tau\sim {\rm DE}(\lambda_{j r}/\sqrt{\phi_{r}\tau})$ for $k=1,\ldots,I_j$,
	where DE($\cdot$) denotes the double-exponential distribution with zero location parameter, we can obtain from  Equations (\ref{APPGZ}) and (\ref{calcf01})  the optimal variational density $q^*(\lambda_{jr})$ for $\lambda_{j r}$ being the generalized inverse Gaussian distribution, i.e.,
	\begin{equation}\label{qlambda}
		q^*(\lambda_{jr})\sim giG(P_{\lambda_{jr}},a_{\lambda_{jr}},b_{\lambda_{jr}}),
	\end{equation}
	where $P_{\lambda_{jr}}=b_{\lambda}-I_j$, $a_{\lambda_{jr}}=2a_{\lambda}$ and $b_{\lambda_{jr}}=2\sum\limits_{k=1}^{I_j}\E^*_{u_{rk}^{(j)}}(|u_{rk}^{(j)}|)E^*_{\phi_{r}}(\phi_{r}^{1/2})E^*_{\tau}(\tau^{1/2})$, in which $\E^*_{u_{rk}^{(j)}}(|u_{rk}^{(j)}|)$ is calculated by the sample absolute moment of $u_{rk}^{(j)}$ based on the updating variational densities of $u_{rk}^{(j)}$, and $E^*_{\tau}(\tau^{1/2})$ is calculated by the sample moment of $\tau^{1/2}$ based on the updating variational densities of $\tau$.

	\subsection{Coordinate Ascent Algorithm for Variational Estimation} \label{section34}
	
	Based on the priors (\ref{proir1})--(\ref{proir3}) and 
	the variational densities (\ref{qu})--(\ref{qlambda}),
	it is easily shown from Equations 
	(\ref{equw4}) and (\ref{MFV01}) that
	the updated ELBO has the form
	\begin{equation}\label{ELBON}
		L\{q(\bbZ), \bxi|m\}=\E^*_{\bbZ}\{\ln h(\bbU,\bxi|m)\}+ \E^*_{\bbZ}\{\ln p(\bbU | \bva)\}+\E^*_{\bbZ}\{\ln p(\bva)\}-\E^*_{\bbZ}\{\ln q(\bbZ)\},	
	\end{equation}
	where $\E^*_{\bbZ}(\cdot)$ represents the expectation taken with respect to the variational density $q^*(\bbZ)$. 
%	\begin{equation}\label{ELBON}
%		\begin{array}{llllll}
%			L\{q(\bbZ), \bxi|m\}
%			&=&\E^*_{\bbU}\{\ln h(\bbU,\bxi|m)\}+\sum\limits_{j=1}^M\sum\limits_{r=1}^R\E^*_{\bu_r^{(j)},\phi_r,\sigma_{jr},\tau}\{\ln p(\bu_r^{(j)}|\tau,\phi_r,\bSi_{jr})\}  \\
%			& & +\E^*_{\tau}\{\ln p(\tau)\}+\E^*_{\Phi}\{\ln p(\bPhi)\}+\sum\limits_{j=1}^M\sum\limits_{r=1}^R\sum\limits_{k=1}^{I_j}\E^*_{\sigma_{jrk},\lambda_{jr}}\{\ln p(\sigma_{jrk}|\lambda_{jr})\} \\
%			& & +\sum\limits_{j=1}^M\sum\limits_{r=1}^R\E^*_{\lambda_{jr}}\{\ln p(\lambda_{jr})\} - \sum\limits_{j=1}^M\sum\limits_{r=1}^R
%			\E^*_{\bu_r^{(j)}}\{\ln q^*(\bu_r^{(j)})\}-\E^*_{\tau}\{\ln q^*(\tau)\} \\
%			& & -\sum\limits_{r=1}^R\E^*_{\phi_r}\{\ln q^*(\phi_r)\}-\sum\limits_{j=1}^M\sum\limits_{r=1}^R\sum\limits_{k=1}^{I_j}\E^*_{\sigma_{jrk}}\{\ln q^*(\sigma_{jrk})\}  \\
%			& & -\sum\limits_{j=1}^M\sum\limits_{r=1}^R\E^*_{\lambda_{jr}}\{\ln q^*(\lambda_{jr})\}.
%		\end{array}
%	\end{equation}
	Note that the above presented ELBO (\ref{ELBON}) involves the variational parameters in $\bxi\in\bbR^{n}$ to be estimated.
	To this end, similarly to \cite{Jaakkola2000BayesianPE}, $\xi_{i}$ ($i=1,2,\ldots,n$) can be calculated by
	\begin{equation}\label{xii}
		\begin{array}{llllll}
			\xi_{i}^2=&{\rm trace}[(\bI_R\otimes \bX_{i(m)}){\rm E}^*_{\bU}\left\lbrace {\rm vec}(\bU^{(-m)}){\rm vec}(\bU^{(-m)\T})\right\rbrace
			(\bI_R\otimes \bX_{i(m)})^{\T}\\
			&
			{\rm E}^*_{\bU}\left\lbrace {\rm vec}(\bU^{(m)}){\rm vec}(\bU^{(m)\T})\right\rbrace].
		\end{array}
	\end{equation}
	More details of the ELBO expression (\ref{ELBON}) and the derivation of the above formula (\ref{xii}) are provided in the Supplementary Material \citep{Jin2025}.
	 Due to the equivalence property of the mode matricization approximations of $\mathcal{W}$ based on the CP decomposition for different modes, the ELBO in equation (\ref{ELBON}) is equal for different $m$. Similarly, variational parameters calculated by equation (\ref{xii}) is equal for different $m$.
		Therefore, $m$ can be a given positive integer from $\{1,2,\ldots,M\}$. In the numerical studies, we set $m=1$ to calculate the ELBO and variational parameters.

	Based on the above discussions, we summarize the detailed steps of the coordinate ascent algorithm for optimizing the updated
ELBO and obtaining the optimal variational density $q^*(\bbZ)$ to approximate the posterior density $p(\bbZ|\by,\bbX,\tau,\bPhi,\bsi,m)$. Consequently,
we can obtain the approximate posterior means or modes of parameters from
their corresponding optimal variational densities. The algorithm is given in the following Algorithm \ref{alg1} with some pre-specified values of $m$ and $R$ as follows.

		\begin{algorithm}[H]
		\caption{Coordinate Ascent Algorithm}\label{alg1}	
		\KwIn{Data$\{(\bmX_i,y_i): i=1,\cdots,n\}$ and $R$ and $m$, where
			$y_{i}\in\{-1,1\}$ and $\bmX_i\in\bbR^{I_{1}\times \cdots \times I_{M}}$.}
		\KwOut{Variational densities $q^*(\bbZ)$ and the posterior modes of parameters $\bbZ$.}
		
		\textbf{Initialization}: %{\color{red}For the given values of $R$ and $m$},
		Take hyperparameters $\alpha_1=\cdots=\alpha_R=1/R\equiv\alpha$, $a_{\lambda}=1$, $b_{\lambda}=a_{\lambda}^{0.5/M}$, $a_{\tau}=\alpha R$, $b_{\tau}=0.5R\sum_{j=1}^MI_j-0.5$; parameters associated with variational densities:
		$\bmu_{\bu_r^{(j)}}=\0$, $\bSi_{\bu_r^{(j)}}=0.1\bI$, $\beta_{\tau}=1$, $b_{r}=0.5\sum_{j=1}^{M}I_j-\alpha$, $a_{jr}=1$, $b_{jrk}=1$, $b_{\lambda_{j r}}=1$ for $k=1,\cdots,I_j$, $j=1,\cdots,M$, and $r=1,\cdots,R$; $\epsilon=10^{-4}$, $t=1$
		and the number of maximum iterations $T=100$.				
		\textbf{Compute} variational densities $q^*(\bu_r^{(j)})$, $q^*(\tau)$, $q^*(\sigma_{j r k})$, $q^*(\phi_{r})$, $q^*(\lambda_{j r})$ via
		Equations (\ref{qu})--(\ref{qlambda}) for $k=1,\cdots,I_j$, $j=1,\cdots,M$, $r=1,\cdots,R$, variational parameter vector $\bxi$ via Equation (\ref{xii}), and $L\{q(\bbZ), \bxi|m\}$ (denoted as $L^{(0)}\{q(\bbZ), \bxi\}$) via Equation (\ref{ELBON}),
		and set $L^{(1)}\{q(\bbZ), \bxi\}=L^{(0)}\{q(\bbZ), \bxi\}+2\epsilon$.
		
		\While{$|L^{(t)}(q(\bbZ), \bxi)-L^{(t-1)}(q(\bbZ), \bxi)|>\epsilon$ and $t<T$}{
			Compute variational density $q^*(\bu_r^{(j)})$ via Equation (\ref{qu}), and update $\bmu_{\bu_r^{(j)}}$ and $\bSi_{\bu_r^{(j)}}$ for $j=1,\cdots,M$, and $r=1,\cdots,R$;\\
			Compute variational density $q^*(\tau)$ via Equation (\ref{qtau}), and update $\E^*_{\tau}(\tau)$, $\E^*_{\tau}(\tau^{-1})$ and $\E^*_{\tau}\{\ln(\tau)\}$;\\
			Compute variational density $q^*(\sigma_{j r k})$ via Equation (\ref{qsigma}), and update $\E^*_{\sigma_{jrk}}(\sigma_{jrk})$,
			$\E^*_{\sigma_{jrk}}(\sigma_{jrk}^{-1})$ and
			$\E^*_{\sigma_{jrk}}\{\ln(\sigma_{jrk})\}$ for $k=1,\cdots,I_j$, $j=1,\cdots,M$, and $r=1,\cdots,R$;\\
			Compute variational density $q^*(\phi_{r})$ via Equation (\ref{qphi}), and updata $\E^*_{\phi_r}\{\ln(\phi_r)\}$, $\E^*_{\phi_r}(\phi_r^{-1})$ and $\E^*_{\phi_r}(\phi_r^{1/2})$ for $r=1,\ldots,R$;\\
			Compute variational density $q^*(\lambda_{j r})$ via Equation (\ref{qlambda}), and updata $\E^*_{\lambda_{jr}}(\lambda_{j r})$, $\E^*_{\lambda_{jr}}(\lambda_{j r}^2)$ and $\E^*_{\lambda_{jr}}\{\ln(\lambda_{j r})\}$ for $j=1,\cdots,M$, $r=1,\cdots,R$;\\
			Compute variational parameter vector $\bxi$ via Equation (\ref{xii});\\
			Compute ELBO $L\{q(\bbZ), \bxi|m\}$ (denoted as $L^{(t)}\{q(\bbZ), \bxi\}$) via Equation (\ref{ELBON}),
			and  $|L^{(t)}\{q(\bbZ), \bxi\}-L^{(t-1)}\{q(\bbZ), \bxi\}|$;\\
			$t\leftarrow t+1$; \\
		}
		\textbf{Return} $q^*(\bu_r^{(j)})$, $q^*(\tau)$, $q^*(\sigma_{j r k})$, $q^*(\phi_{r})$, $q^*(\lambda_{j r})$ for $k=1,\cdots,I_j$, $j=1,\cdots,M$, $r=1,\cdots,R$, and posterior modes of parameters $\bbZ$ corresponding to maximizer of ELBOs.
	\end{algorithm}

	In general,  $\epsilon$ is taken as $\epsilon=10^{-4}, 10^{-5}$ or $10^{-6}$. Different values of $\epsilon$ lead to different convergence rates of the algorithm presented above. Generally, the smaller $\epsilon$ is, the slower the convergence of the algorithm is. According to our experience, the choice of a smaller or moderately small $\epsilon$ has little effect on variational Bayesian estimations of parameters. Thus, we set $\epsilon=10^{-4}$ in the simulation studies and real example analysis. Similarly to the Gibbs sampler, the order of the steps presented above for optimizing the ELBO value has little effect on variational Bayesian estimates of parameters.
	Note that the choice of $R$ is of critical importance. Generally, the problem of selecting $R$ can be
	regarded as the problem of model selection. Thus, similarly to the idea of model selection,
	$R$ can be selected by minimizing Akaike information criterion (AIC) or Bayesian information criterion (BIC)
	or maximizing the ELBO \citep{ZhangYang2024}.
	In simulation studies and real example analysis conducted below, we choose an appropriate $R$ via maximizing the ELBO.

	\subsection{Classification Prediction} \label{section35}
	
%	By the argument of Section \ref{section34}, when the above introduced coordinate ascent algorithm converges, we can obtain the optimal variational density $q^*(\bbZ)$ for approximating the posterior density $p(\bbZ|\by,\bbX,\tau,\bPhi,\bsi,m)$. Consequently,
%	we can obtain the approximate posterior means or modes of parameters from
%	their corresponding optimal variational densities. The algorithm for computing variational Bayesian estimates of parameters
%	in the considered model is given in the following Algorithm \ref{alg1}. 	

	Based on the optimal variational density $q^*(\bbU)$ and variational Bayesian estimates of parameters computed with Algorithm \ref{alg1},
	one can compute the probability of the class $Y=1$ for a given tensor $\bmX$ with the following form:
	\begin{equation}\label{PRDC}
		\begin{array}{llll}
			\Pr(Y=1 \mid \bmX, \bbD) &=& \dfrac{1}{J}\sum\limits_{j=1}^{J} g({\rm trace}\{\bW_{(1)}^{(j)}\bX_{(1)}^{\top}\})\\
			&\approx& \dfrac{1}{J}\sum\limits_{j=1}^{J} g\left\{{\rm vec}(\bU_{(j)}^{(1)})^{\top}
			(\bI_R\otimes\bX_{(1)})\textmd{vec}(\bU_{(j)}^{(-1)})\right\},
		\end{array}
	\end{equation}
	\noindent where $\bW_{(1)}$ and $\bX_{(1)}$ are the mode-1 matricization versions of tensors $\bmW$ and $\bmX$, respectively,
	$\{\bW_{(1)}^{(1)},\ldots,\bW_{(1)}^{(J)}\}$ is constructed based on
	%	the sample drawn from the posterior distribution $p(\bW_{(1)}|\bbD)$;
	$\{\bU_{(1)}^{(1)},\bU_{(1)}^{(-1)},\ldots,\bU_{(J)}^{(1)},\bU_{(J)}^{(-1)}\}$ which is the sample drawn from the optimal variational density $q^*(\bbU)$,  and $J=1000$ in the numerical studies.
	Thus, we can classify an individual associated with tensor $\bmX$ as some specific class via the following criterion:
	an individual associated with tensor $\bmX$ is classified as label $1$ if its class probability $\Pr(y=1 \mid \bmX, \bbD)$ is larger than the pre-specified threshold (e.g., $a$); and otherwise, it is identified as label $-1$.
	Based on the above argument,  we can summarize the algorithm for forecasting classification as the following Algorithm \ref{alg2}.

	\begin{algorithm}[H]
		% \SetAlgoNoLine  %去掉之前的竖线
		\caption{Forecasting Classification}\label{alg2}			
		\KwIn{Training data $\{(\bmX_i,y_i): i=1,\cdots,n\}$, testing data $\bmX_0$ and threshold $a$, where
			$y_{i}\in\{-1,1\}$ and $\bmX_i\in\bbR^{I_{1}\times \cdots \times I_{M}}$ and $\bmX_0\in\bbR^{I_{1}\times \cdots \times I_{M}}$.}
		\KwOut{Forecasting label $y$ associated with tensor $\bmX_0$ in testing data.}
		\textbf{Implement Algorithm 1 and return} the optimal variational densities $q^*(\bu_r^{(j)})$, $q^*(\tau)$, $q^*(\sigma_{j r k})$, $q^*(\phi_{r})$, $q^*(\lambda_{j r})$ for $k=1,\cdots,I_j$, $j=1,\cdots,M$, $r=1,\cdots,R$, and the posterior modes of unknown parameters $\bbZ$.\\
		\textbf{Sample} from the optimal variational densities $q^*(\bu_r^{(j)})$ and construct the factor matrices $\{\bU_{(1)}^{(1)},\bU_{(1)}^{(-1)},\ldots,\bU_{(J)}^{(1)},\bU_{(J)}^{(-1)}\}$.\\
		\textbf{Calculate}: Give the tensor $\bmX_0$ in testing data, compute
		the class probability $\Pr(Y=1 \mid \bmX, \bbD)$ via Equation (\ref{PRDC}). \\
		\textbf{Classification}: For the threshold $a$, if $\Pr(Y=1 \mid \bmX, \bbD)>a$,
		classify the corresponding individual as label $1$, and $-1$ otherwise.
	\end{algorithm}

	 Note that there are many methods and criteria for selecting an optimal threshold $a$, for example, efficiency, misclassification-cost, odds ratio, the kappa index \citep{GREINER200023} and Youden index \citep{Fluss2005}.
		In numerical studies below, we select the optimal threshold $a$ by maximizing the Youden index, which has the virtue of being the easiest to implement and does not require further information such as prevalence rates and decision error costs \citep{Fluss2005}. The details are as follows.
		Based on several threshold values of $a$ in the set of alternative values taken from the interval $(0,1)$, we first obtain the forecasting labels of training data via Algorithm 2, and then calculate sensitivities and specificities (denoted as Sensitivity($a$) and Specificity($a$))
		defined below, select the optimal threshold $a^*$ by maximizing the Youden index, i.e., $a^*=\arg\max_a\{{\rm Sensitivity}(a) + {\rm Specificity}(a) - 1\}$.

	\section{Simulation Studies}\label{Simulation}

	In this section, simulation studies are conducted to investigate the finite sample performance of the proposed method (denoted as `VBLTR').
	For comparison, we also consider the following seven methods: Bayesian tensor logistic regression (denoted as `BTLR') \citep{Wu2022Tang}, C-support vector classification (C-SVC) \citep{Boser1992,Cortes1995}, $\nu$-support vector regression ($\nu$-SVR) \citep{Scholkopf2000}, linear-support vector regression (L-SVR) \citep{Ho2012},
	K-nearest neighbors algorithm (KNN), random forests (RF),  and Convolutional Neural Network (CNN), which are implemented with \textit{libsvm} library \citep{Chang2011} and random forests package \citep{breiman2001random}.
	Note that the proposed method adopts the hyperparameters and initialization of parameters given in Section \ref{section34},
	 and chooses the appropriate $R$ associated with the CP decomposition via maximizing the ELBO in Section \ref{section34}. Additional simulation studies for assessing the robustness of the proposed method are provided in the Supplementary Material \citep{Jin2025}.

	In the simulation study, we generate $n_1$ observations $\{\bmX_i,y_i\}_{i=1}^{n_1}$ in which the components of the $I_1\times I_2\times I_3$ tensor $\bmX_i$ are independently sampled from the normal distribution with mean $\mu_1=0$ and variance $1$,
	and generate $n_2$ observations $\{\bmX_i,y_i\}_{i=n_1+1}^{n_1+n_2}$ in which the components of the $I_1\times I_2\times I_3$ tensor $\bmX_i$ are independently drawn from the normal distribution with mean $\mu_2$ and variance $1$,
	where $y_i$ for $i=1,2,\ldots, n_1+n_2$ is the observation of binary response
	with the possible values being $\{-1,1\}$ sampled from the Bernoulli distribution with  possibility
	$\Pr\left(Y=y_{i} \mid \bmX_{i}, \bmW\right)
	=1/(1+\exp\left\{-y_{i}\left\langle\bmW, \bmX_{i}\right\rangle\right\})$.
	We set the components $\bmW(i,j,k)=1$ for $i=1,\ldots,4; j=2,\ldots,5;\\ k=1,2,3$ and others being zero in the coefficient tensor $\bmW\in\bbR^{I_1\times I_2\times I_3}$.
	Thus, the total sample size is $N=n_1+n_2$. The datasets $\{\bmX_i,y_i\}_{i=1}^{N}$ are randomly divided into training data of $80\%$ for constructing classifier and testing data of $20\%$ for assessing the performance of the constructed classifier. We repeated the experiment 100 times.

To assess the performance of VBLTR and BTLR methods in estimating tensor coefficient, we consider the mean absolute error and root mean square error, i.e.,
$
{\rm MAE}=\sum_{j=1}^{100}\|\bfm{\mathcal{W}}^{(0)}-\widehat{\bfm{\mathcal{W}}}^{(j)}\|_1/100\quad {\rm and}\quad {\rm RMSE}=\sqrt{\sum_{j=1}^{100}\|\bfm{\mathcal{W}}^{(0)}-\widehat{\bfm{\mathcal{W}}}^j\|_2^2/100},$
where $\|\cdot\|_1$ denotes the sum of absolute values of tensor elements,  $\|\cdot\|_2^2$ are the sum of squares of tensor elements, and  $\bfm{\mathcal{W}}^{(0)}$ and $\widehat{\bfm{\mathcal{W}}}^{(j)}$ is the true value and estimator  of tensor coefficient $\bfm{\mathcal{W}}$ at the $j$th repetition, respectively. The smaller both metrics are, the better the performance of the estimation method is.
		To assess the performance of the identifiability of active components, we calculate the proportions of truly active components correctly identified as active ones (denoted as `TAR'), truly inactive components incorrectly identified as active ones (denoted as  `FAR'),
		components correctly identified (denoted as `Rate').
		Closer to one the TAR and Rate values are or closer to zero the FAR is, the better the performance of component selection method is.
		Moreover, we also show the computational times in seconds for the VBLTR and BTLR methods for comparing their computing cost (denoted as `CTime'). These calculations are done on a laptop with an Intel(R) Xeon(R) W-11955M CPU @2.60GHz and 64 GB memory.

	To assess the classification performance of the constructed classifier, we consider the following five metrics:
	sensitivity, specificity, accuracy, precision and F1-score, which are defined as
	$${\rm Sensitivity=\frac{TP}{TP+FN},	Specificity=\frac{TN}{TN+FP},
		Accuracy=\frac{TP+TN}{TP+TN+FP+FN}},$$
	$${\rm Precision=\frac{TP}{TP +FP},	F1-score=\frac{\rm 2\times TP}{2 \times TP+FP+FN}},$$
	respectively, where TP is the number of true positive subjects, TN is the number of true negative subjects,
	FP is the number of false positive subjects, and FN is the number of false negative subjects,
	TP+FN is the number of actual positive subjects,
	FP+TN is the number of actual negative subjects, TP+FP is the number of positive predictions,
	and FN+TN is the number of negative predictions.
	Sensitivity is the possibility that a classifier correctly identifies true positive subjects,
	specificity is the possibility that a classifier correctly detects true negative subjects,
	accuracy is the proportion of correctly identifying true positive and true negative subjects,
	precision is the proportion of true positives among the predicted positive subjects,
	and F1-score is the harmonic mean of the precision and sensitivity.
	Also, we consider the areas under the curve of receiver operating characteristic (AUC) as an evaluation metric
	for the assessment of a classifier.
	Closer to 100\% these metrics are, the better the performance of the classifier is.

	\begin{table}[H]
		%\centering
		\fontsize{7}{8}\selectfont
		\begin{center}
			\caption{Means (standard deviations) of MAE, RMSE, TAR, FAR and Rate values for coefficient tensor estimates in Experiment 1.}
			\label{sampleestimator}
			\vspace{0.2cm}
			\renewcommand{\arraystretch}{1.3} \tabcolsep 0.05in %\doublerulesep 1.5pt   NN
			\begin{tabular}{cccccccc}\hline  \hline
				$N$&Method&MAE&RMSE&TAR&FAR&Rate&CTime (s) \\ \hline			
				100&VBLTR& {\bf 0.046}(0.003)	 &{\bf  0.183}(0.010)	 &0.905(0.332)	 &{\bf 0.040}(0.016)	 &{\bf 0.946}(0.017) & 750 \\
				&BTLR&3.135(3.061)	 &7.227(7.723)	 &0.916(0.164)	 &0.799(0.197)	 &0.229(0.186) & 40640 \\
				300&VBLTR& {\bf 0.031}(0.004)	 &{\bf  0.101}(0.014)	 &1.000(0.000)	 &{\bf 0.085}(0.024)	 &{\bf 0.918}(0.023) & 1750\\
				&BTLR&2.989(2.457)	 &7.846(6.119)	 &1.000(0.000)	 &0.659(0.227)	 &0.367(0.217) & 47656\\
				500&VBLTR& {\bf 0.023}(0.004)	 &{\bf  0.062}(0.014)	 &1.000(0.000)	 &{\bf 0.095}(0.029)	 &{\bf 0.909}(0.027) & 3150\\
				&BTLR&0.201(0.457)	 &0.626(1.484)	 &1.000(0.000)	 &0.213(0.138)	 &0.795(0.133) & 58488\\
				1000&VBLTR& {\bf 0.016}(0.004)	 &{\bf 0.040}(0.010)	 &1.000(0.000)	 &{\bf 0.043}(0.020)	 &{\bf 0.959}(0.019) & 4050\\
				&BTLR&0.066(0.087)	 &0.220(0.325)	 &1.000(0.000)	 &0.145(0.059)	 &0.861(0.057) & 85652\\
				\hline  \hline
			\end{tabular}
		\end{center}
	\end{table}

	\textbf{Experiment 1}. To assess the effect of the sample sizes, we considers the following four sample sizes: $(n_1,n_2)=(20,80),(60,240),(100,400),(200,800)$, i.e., $N=100,300,500,1000$ corresponding to small, moderate and large sample sizes, respectively.
	In this experiment, we set $(I_1,I_2,I_3)=(10,12,10)$ and $\mu_2=0.2$. Results for tensor coefficient estimates and six metrics of eight classifiers are presented in Tables \ref{sampleestimator} and \ref{samplesize}, respectively.

		Inspection of Table \ref{sampleestimator} shows that (i) the proposed method (i.e., VBLTR) behaves better than the BTLR method in estimating coefficient tensor in that the former has smaller MAE and RMSE values than the latter; (ii) the proposed method can better distinguish active variables from inactive ones than the BTLR method in that the Rate values of the former are closer to one than those of the latter and  the FAR values of the former are closer to zero than those of the latter even if their TAR values are almost identical; (iii) the proposed method demands significantly less computational cost than the BTLR method, which imply that the proposed method is more efficient than the BTLR method \citep{Wu2022Tang}.

	\begin{table}[H]
		%\centering
		\fontsize{7}{8}\selectfont
		\begin{center}
			\caption{Means (standard deviations) of six metrics (\%) for eight classifiers in Experiment 1.}
			\label{samplesize}
			%				\vspace{0.01cm}
			\renewcommand{\arraystretch}{1.3} \tabcolsep 0.08in %\doublerulesep 1.5pt   NN
			\begin{tabular}{cccccccc}\hline  \hline
				$N$&Method&Sensitivity&Specificity&AUC&Accuracy&Precision&F1-score \\ \hline			
				100&VBLTR&87.68(0.09)	 &{\bf  60.24}(0.33)	 &{\bf 84.11}(0.12)	 &{\bf 83.35}(0.09)	 &{\bf 91.77}(0.07)	 &89.36(0.06)  \\
				&BTLR&81.54(0.11)	 &44.81(0.29)	 &65.19(0.21)	 &74.85(0.11)	 &87.14(0.08)	 &83.76(0.08)		\\
				&C-SVC&  99.82(0.01)	 &5.25(0.19)	 &59.89(0.28)	 &83.15(0.08)	 &83.24(0.08)	 &90.57(0.05)	 \\
				&$\nu$-SVR& {\bf 100.00}(0.00)	 &0.00(0.00)	 &73.81(0.17)	 &82.42(0.07)	 &82.42(0.07)	 &{\bf 90.19}(0.04)   \\
				&L-SVR&  {\bf 100.00}(0.00)	 &0.00(0.00)	 &72.30(0.17)	 &82.42(0.07)	 &82.42(0.07)	 &{\bf 90.19}(0.04)  \\
				&KNN& 99.00(0.03)	 &5.60(0.16)	 &67.81(0.17)	 &82.47(0.08)	 &83.04(0.07)	 &90.12(0.05)	   \\
				&RF& 96.62(0.05)	 &4.71(0.13)	 &53.86(0.19)	 &80.36(0.08)	 &82.63(0.08)	 &88.82(0.05)  \\
				&CNN& 99.94(0.01)	 &0.00(0.00)	 &61.29(0.18)	 &82.37(0.07)	 &82.41(0.07)	 &90.16(0.04)  \\		
				300&VBLTR&95.01(0.04)	 &{\bf 82.34}(0.13)	 &{\bf 97.35}(0.02)	 &{\bf 92.63}(0.04)	 &{\bf 96.04}(0.03)	 &{\bf 95.45}(0.02)   \\
				&BTLR&93.82(0.04)	 &72.97(0.16)	 &89.10(0.08)	 &90.12(0.04)	 &94.18(0.03)	 &93.93(0.02)	\\
				&C-SVC&  96.64(0.03)	 &25.37(0.19)	 &78.52(0.08)	 &84.12(0.04)	 &85.95(0.05)	 &90.87(0.03)\\
				&$\nu$-SVR&	{\bf 100.00}(0.00)	 & 0.00(0.00)	 & 78.51(0.08)	 & 82.37(0.04)	 & 82.37(0.04)	 & 90.27(0.03)   \\
				&L-SVR&  {\bf 100.00}(0.00)	 & 0.00(0.00)	 & 74.07(0.09)	 & 82.37(0.04)	 & 82.37(0.04)	 & 90.27(0.03)  \\
				&kNN&  98.70(0.02)	 & 7.67(0.11)	 & 66.83(0.11)	 & 82.68(0.04)	 & 83.35(0.04)	 & 90.31(0.03)  \\
				&RF&  97.00(0.03)	 & 3.56(0.06)	 & 54.12(0.10)	 & 80.55(0.05)	 & 82.47(0.04)	 & 89.07(0.03)  \\
				&CNN&  93.35(0.09)	 &12.20(0.14)	 &61.61(0.11)	 &79.22(0.08)	 &83.32(0.04)	 &87.79(0.05)	  \\		
				500&VBLTR&94.95(0.03)	 &{\bf 86.67}(0.09)	 &{\bf 98.12}(0.01)	 &{\bf 93.49}(0.03)	 &{\bf 97.18}(0.02)	 &{\bf 96.01}(0.02) \\
				&BTLR&94.93(0.03)	 &85.79(0.10)	 &97.48(0.02)	 &93.28(0.03)	 &96.94(0.02)	 &95.88(0.02)		\\
				&C-SVC&  95.47(0.03)	 &32.60(0.16)	 &81.53(0.07)	 &84.85(0.04)	 &87.51(0.04)	 &91.24(0.02) \\
				&$\nu$-SVR& {\bf 100.00}(0.00)	 & 0.00(0.00)	 & 81.52(0.07)	 & 83.09(0.04)	 & 83.09(0.04)	 & 90.72(0.02)	   \\
				&L-SVR& 99.95(0.00)	 & 0.88(0.03)	 & 76.41(0.07)	 & 83.20(0.04)	 & 83.21(0.04)	 & 90.77(0.02)   \\
				&KNN&  98.50(0.02)	 & 8.68(0.08)	 & 67.59(0.08)	 & 83.32(0.04)	 & 84.13(0.04)	 & 90.71(0.02)  \\
				&RF&  97.56(0.02)	 & 4.45(0.06)	 & 55.65(0.08)	 & 81.82(0.04)	 & 83.38(0.04)	 & 89.86(0.02)  \\
				&CNN& 91.39(0.07)	 &26.80(0.18)	 &69.59(0.10)	 &80.53(0.06)	 &86.10(0.04)	 &88.48(0.04)	 \\
				1000&VBLTR&94.78(0.02)	 &{\bf 91.12}(0.05)	 &{\bf 98.47}(0.01)	 &{\bf 94.16}(0.02)	 &{\bf 98.10}(0.01)	 &{\bf 96.39}(0.01) \\
				&BTLR&94.59(0.02)	 &90.38(0.05)	 &98.35(0.01)	 &93.86(0.02)	 &97.94(0.01)	 &96.22(0.01)	 	\\
				&C-SVC&  93.92(0.02)	 &42.95(0.10)	 &82.91(0.04)	 &85.16(0.02)	 &88.81(0.02)	 &91.26(0.01)	 \\
				&$\nu$-SVR&	{\bf 99.93}(0.00)	 & 1.07(0.03)	 & 82.90(0.04)	 & 82.88(0.02)	 & 82.90(0.03)	 & 90.60(0.01)   \\
				&L-SVR&  98.21(0.01)	 &11.00(0.06)	 &72.89(0.06)	 &83.18(0.03)	 &84.12(0.02)	 &90.60(0.02)	    \\
				&KNN&  98.31(0.01)	 & 10.73(0.07)	 & 68.62(0.05)	 & 83.20(0.02)	 & 84.09(0.03)	 & 90.62(0.02)  \\
				&RF& 97.36(0.01)	 & 4.99(0.04)	 & 56.18(0.06)	 & 81.43(0.03)	 & 83.11(0.02)	 & 89.65(0.02)     \\
				&CNN& 91.09(0.04)	 &46.29(0.13)	 &79.14(0.06)	 &83.37(0.04)	 &89.10(0.03)	 &90.00(0.02)	  \\
				\hline  \hline
			\end{tabular}
		\end{center}
	\end{table}

	Examination of Table \ref{samplesize} % and Figure \ref{simuFig5}
	shows that
	(i)	the proposed classifier outperforms other classifiers in that specificity, AUC, accuracy and precision values of the former are larger than those of the latter regardless of the sample size, and F1-score values of the former are larger than those of the latter except for $N=100$;	(ii) although sensitivity and F1-score values of $\nu$-SVR and L-SVR methods are larger than those of other competing methods for $N=100$, but specificity values of $\nu$-SVR and L-SVR methods are zero, which implies that $\nu$-SVR and L-SVR methods behave unsatisfactorily; (iii) the proposed classifier consistently outperforms the BTLR classifier in that six metrics of the former are larger than those of the latter regardless of the sample sizes; (iv) six metrics of the proposed method increase as the sample size increases, while other classifiers have unstable values of metrics. The above findings imply that the proposed classifier behaves well regardless of the sample sizes. %To better compare the performance of the %proposed classifier and seven competing classifiers, means of six metrics for eight classifiers with four sample sizes: $N=100$, 300, 500 and 1000 %are presented in Figure \ref{simuFig00}. Figure \ref{simuFig00} intuitively shows that the proposed classifier has larger metrics than other %classifiers except for sensitivity regardless of the sample sizes.

	%\begin{figure}[H]
	%		\centering
	%		\scalebox{0.18}[0.2]{\includegraphics{differentsamplesize.png}}\vspace{-1mm}
	%\caption{\footnotesize{Means of six metrics among 100 replications for eight classifiers with (a) $N=100$, (b) $N=300$, (c) $N=500$ and (d) $N=1000$	%in Experiment 1.}}
	%		\label{simuFig00}
	%	\end{figure}

\begin{table}[H]
	%\centering
	\fontsize{7}{8}\selectfont
	\begin{center}
		\caption{Means (standard deviations) of MAE, RMSE, TAR, FAR and Rate values for coefficient tensor estimates in Experiment 2.}
		\label{dimensestimator}
		%				\vspace{0.2cm}
		\renewcommand{\arraystretch}{1.3} \tabcolsep 0.035in %\doublerulesep 1.5pt   NN
		\begin{tabular}{cccccccc}\hline  \hline
			$(I_1,I_2,I_3)$&Method&MAE&RMSE&TAR&FAR&Rate&CTime(s) \\ \hline			
			(5,6,5)&VBLTR&{\bf 0.057}(0.017)	 &{\bf 0.085}(0.027)	 &1.000(0.000)	 &{\bf 0.122}(0.089)	 &{\bf 0.917}(0.061) & 1800  \\
			&BTLR&0.209(0.299)	 &0.333(0.505)	 &1.000(0.000)	 &0.338(0.165)	 &0.770(0.112) & 55096 \\
			(10,12,10)&VBLTR& {\bf 0.016}(0.004)	 &{\bf 0.040}(0.010)	 &1.000(0.000)	 &{\bf 0.043}(0.020)	 &{\bf 0.959}(0.019) & 4050  \\
			&BTLR& 0.066(0.087)	 &0.220(0.325)	 &1.000(0.000)	 &0.145(0.059)	 &0.861(0.057) & 85652\\
			(15,18,15)&VBLTR&{\bf 0.008}(0.001)	 &{\bf 0.027}(0.004)	 &1.000(0.000)	 &{\bf 0.021}(0.006)	 &{\bf 0.980}(0.006) & 11050 \\
			&BTLR&0.061(0.089)	 &0.300(0.476)&1.000(0.000)	 &0.106(0.067)	 &0.895(0.066)&174762\\
			(20,24,20)&VBLTR&{\bf 0.004}(0.001)	 &{\bf 0.019}(0.003)	 &{\bf1.000}(0.000)	 &{\bf 0.011}(0.003)	 &{\bf 0.989}(0.003) & 39750\\
			&BTLR&0.079(0.080)	 &0.495(0.511)	 &0.949(0.209)	 &0.116(0.079)	 &0.884(0.079) &316335\\
			\hline  \hline
		\end{tabular}
	\end{center}
\end{table}

\textbf{Experiment 2}. To assess the effect of dimension of mode, we consider the following cases:
$(I_1,I_2,I_3)=(5,6,5), (10,12,10), (15,18,15)$ and $(20,24,20)$ corresponding to low and moderate dimensions, respectively. In this experiment, we
take $(n_1,n_2)=(200,800)$ and $\mu_2=0.2$. Results for tensor coefficient estimates and six metrics are presented in Tables \ref{dimensestimator} and \ref{dimension}. % and Figure \ref{Fig7}. 		
  From Table \ref{dimensestimator}, we observe that the MAE, RMSE, FAR and CTime values of the proposed method are smaller than those of the BTLR method, and the TAR and Rate values of the proposed method are closer to one than those of the BTLR method regardless of the dimensions, which imply that the proposed method behaves better than the BTLR method in estimating coefficient tensor and identifying active variables. % and the performance of the proposed method is robust for various dimensions of modes.
	More importantly, the computational burden of the proposed method is typically much smaller than that of the BTLR method.

 Inspection of Table \ref{dimension} %and Figure \ref{Fig7}
	indicates that (i) the proposed classifier has the same performance as the BTLR method for the low dimensions of mode (e.g., $(I_1,I_2,I_3)=(5,6,5)$ and $(10,12,10)$), while the proposed method outperforms the BTLR method for the relatively moderate dimensions (e.g., $(I_1,I_2,I_3)=(20,24,20)$), which implies that the proposed classifier is efficient in dealing with high-dimensional tensor issue;
	(ii) the proposed classifier has the similar performance as other six competing classifiers in terms of sensitivity and F1-score, but consistently behaves better than other six competing classifiers in terms of specificity, AUC, accuracy and precision values;
	(iii) the C-SVC, $\nu$-SVR, L-SVR, KNN, RF and CNN classifiers behave unstable and suffer from negligible effect of the dimensions of modes
	in terms of specificity, AUC, accuracy, precision and F1-score.

\begin{table}[H]
	%\centering
	\fontsize{7}{8}\selectfont
	\begin{center}
		\caption{Means (standard deviations) of six metrics (\%) for eight classifiers in Experiment 2.}
		\label{dimension}
		\vspace{0.2cm}
		\renewcommand{\arraystretch}{1.3} \tabcolsep 0.05in %\doublerulesep 1.5pt   NN
		\begin{tabular}{cccccccc}\hline  \hline
			$(I_1,I_2,I_3)$&Method&Sensitivity&Specificity&AUC&Accuracy&Precision&F1-score \\ \hline			
			(5,6,5)&VBLTR&94.33(0.02)	 &{\bf 93.49}(0.05)	 &{\bf 98.76}(0.01)	 &94.18(0.02)	 &{\bf 98.58}(0.01)	 &96.39(0.01)	  \\
			&BTLR& 94.47(0.02)	 &93.41(0.05)	 &98.72(0.01)	 &{\bf 94.28}(0.02)	 &98.56(0.01)	 &{\bf 96.45}(0.01) \\
			&C-SVC& 95.21(0.02)	 &67.17(0.09)	 &94.45(0.02)	 &90.37(0.02)	 &93.29(0.02)	 &94.22(0.01)	    \\
			&$\nu$-SVR& {\bf 99.30}(0.01)	 & 27.88(0.16)	 & 93.81(0.02)	 & 86.86(0.03)	 & 86.82(0.03)	 & 92.59(0.02) \\
			&L-SVR& 84.41(0.04)	 &83.27(0.07)	 &92.31(0.03)	 &84.24(0.03)	 &96.05(0.02)	 &89.78(0.03)	 \\
			&KNN& 98.98(0.01)	 & 8.93(0.07)	 & 73.29(0.05)	 & 83.35(0.02)	 & 83.82(0.02)	 & 90.75(0.01)  \\
			&RF& 97.04(0.01)	 & 8.54(0.05)	 & 63.20(0.06)	 & 81.69(0.03)	 & 83.49(0.02)	 & 89.73(0.02) \\
			&CNN& 90.83(0.04)	 &64.56(0.11)	 &88.20(0.04)	 &86.24(0.03)	 &92.44(0.02)	 &91.56(0.02)	  \\		
			(10,12,10)	&VBLTR&94.78(0.02)	 &{\bf 91.12}(0.05)	 &{\bf 98.47}(0.01)	 &{\bf 94.16}(0.02)	 &{\bf 98.10}(0.01)	 &{\bf 96.39}(0.01)   \\
			&BTLR& 94.59(0.02)	 &90.38(0.05)	 &98.35(0.01)	 &93.86(0.02)	 &97.94(0.01)	 &96.22(0.01)	\\
			&C-SVC& 93.92(0.02)	 &42.95(0.10)	 &82.91(0.04)	 &85.16(0.02)	 &88.81(0.02)	 &91.26(0.01) \\
			&$\nu$-SVR& {\bf 99.93}(0.00)	 & 1.07(0.03)	 & 82.90(0.04)	 & 82.88(0.02)	 & 82.90(0.03)	 & 90.60(0.01)\\
			&L-SVR& 98.21(0.01)	 &11.00(0.06)	 &72.89(0.06)	 &83.18(0.03)	 &84.12(0.02)	 &90.60(0.02) \\
			&kNN& 98.31(0.01)	 & 10.73(0.07)	 & 68.62(0.05)	 & 83.20(0.02)	 & 84.09(0.03)	 & 90.62(0.02)  \\
			&RF& 97.36(0.01)	 & 4.99(0.04)	 & 56.18(0.06)	 & 81.43(0.03)	 & 83.11(0.02)	 & 89.65(0.02) \\
			&CNN& 91.09(0.04)	 &46.29(0.13)	 &79.14(0.06)	 &83.37(0.04)	 &89.10(0.03)	 &90.00(0.02)	 \\		
			(15,18,15)&VBLTR&94.73(0.02)	 &{\bf 88.69}(0.07)	 &{\bf 98.28}(0.01)	 &{\bf 93.71}(0.02)	 &{\bf 97.66}(0.01)	 &{\bf96.15}(0.01) \\
			&BTLR&94.76(0.02)	 &87.40(0.07)	 &97.90(0.01)	 &93.55(0.02)	 &97.43(0.01)	 &96.06(0.01)\\
			&C-SVC& 93.13(0.03)	 &40.56(0.11)	 &78.53(0.04)	 &84.26(0.02)	 &88.58(0.03)	 &90.74(0.01)	  \\
			&$\nu$-SVR& {\bf 100.00}(0.00)	 & 0.00(0.00)	 & 78.53(0.04)	 & 83.12(0.03)	 & 83.12(0.03)	 & 90.76(0.02) \\
			&L-SVR& 99.99(0.00)	 &0.06(0.00)	 &76.74(0.04)	 &83.12(0.03)	 &83.13(0.03)	 &90.76(0.02)	 \\
			&KNN& 96.22(0.02)	 & 18.09(0.10)	 & 70.17(0.05)	 & 83.05(0.02)	 & 85.30(0.03)	 & 90.39(0.02) \\
			&RF& 97.27(0.01)	 & 4.39(0.03)	 & 55.14(0.05)	 & 81.58(0.03)	 & 83.36(0.03)	 & 89.75(0.02) \\
			&CNN& 88.31(0.05)	 &34.76(0.13)	 &67.81(0.07)	 &79.25(0.04)	 &87.01(0.03)	 &87.54(0.03)	   \\
			(20,24,20)&VBLTR&94.68(0.02)	 &{\bf 88.56}(0.06)	 &{\bf 98.23}(0.01)	 &{\bf 93.64}(0.02)	 &{\bf 97.58}(0.01)	 &{\bf 96.09}(0.01)	\\
			&BTLR&93.35(0.06)	 &82.70(0.08)	 &95.36(0.06)	 &91.52(0.06)	 &96.28(0.02)	 &94.69(0.04)	\\
			&C-SVC& 92.17(0.04)	 &44.87(0.16)	 &77.73(0.05)	 &84.10(0.03)	 &89.14(0.03)	 &90.54(0.02)	 \\
			&$\nu$-SVR&	{\bf 100.00}(0.00)	 & 0.00(0.00)	 & 77.74(0.05)	 & 82.96(0.03)	 & 82.96(0.03)	 & 90.66(0.02)   \\
			&L-SVR& {\bf 100.00}(0.00)	 &0.00(0.00)	 &77.52(0.06)	 &82.96(0.03)	 &82.96(0.03)	 &90.66(0.02)	   \\
			&KNN& 94.97(0.03)	 & 26.71(0.13)	 & 72.07(0.06)	 & 83.36(0.03)	 & 86.39(0.03)	 & 90.42(0.02)  \\
			&RF& 97.47(0.01)	 & 4.17(0.03)	 & 54.65(0.05)	 & 81.57(0.03)	 & 83.20(0.03)	 & 89.74(0.02) \\
			&CNN& 87.21(0.07)	 &33.51(0.14)	 &65.09(0.08)	 &78.10(0.05)	 &86.59(0.03)	 &86.72(0.03)	 \\
			\hline  \hline
		\end{tabular}
	\end{center}
\end{table}

%		\begin{figure}[H]
	%			\centering
	%			\scalebox{0.18}[0.22]{\includegraphics{differentdimension.png}}\vspace{-3mm}
	%			\caption{\footnotesize{Means of six metrics among 100 replications for eight classifiers with different dimensions of modes}}
	%			\label{Fig7}
	%		\end{figure}

\section{Real examples}

In this section, the flower image data and the validated Chest X-ray image data are used to illustrate the proposed methodologies. The results of the validated Chest X-ray image data analysis for investigating order-2 tensor data can be found in the Supplementary Material \citep{Jin2025}.
For comparison, we also consider the BTLR, C-SVC, $\nu$-SVR, L-SVR, KNN, RF and CNN classifiers.

%\textbf{(1) Flower Image Data}

\begin{figure}[H]
	\centering
	\scalebox{0.4}[0.4]{\includegraphics{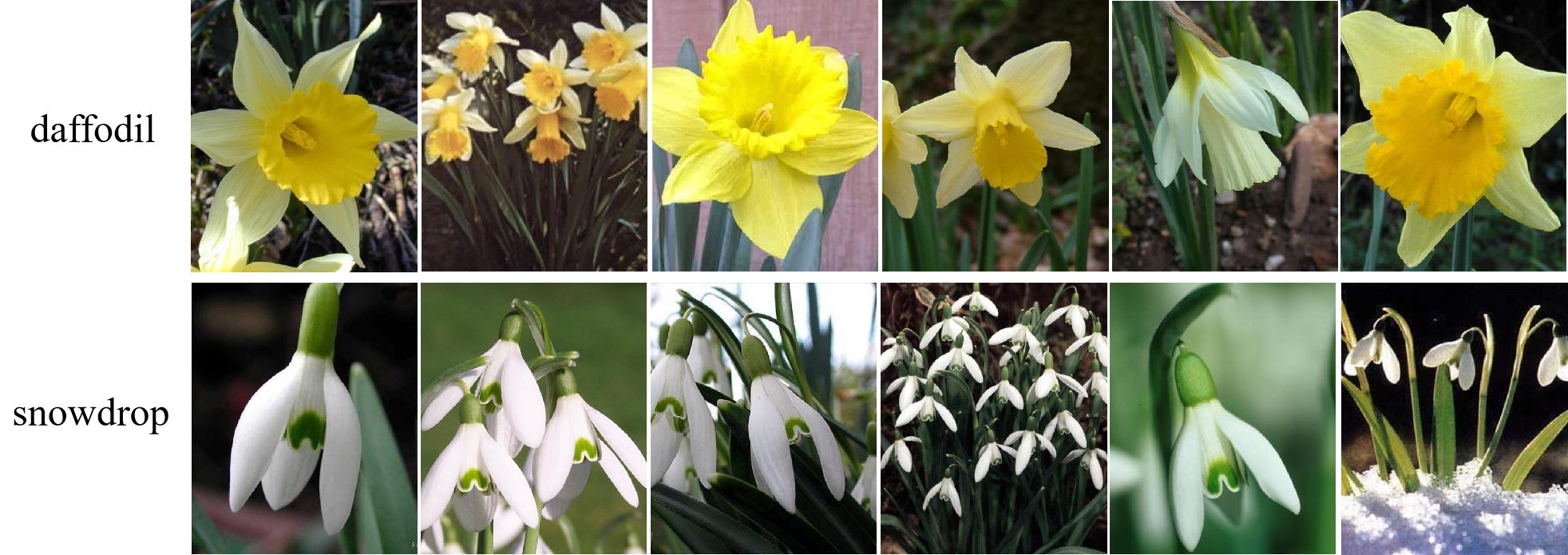}}\vspace{-3mm}
	\caption{\footnotesize{Some examples for snowdrop and daffodil flowers}}
	\label{Eflowers}
\end{figure}

We consider the 17 category flower dataset in the United Kingdom, which is created by Maria-Elena Nilsback and Andrew Zisserman and can be obtained from the website: https://www.robots.ox.ac.uk/$\sim$vgg/data/flowers/17/. Each class has 80 images with different
poses, sizes and perspectives. Flowers includes sunflowers, snowdrop, daffodil, hyacinths and chrysanthemums, and among others.
The dataset has been analyzed in the flower recognition literature, and is one of the most representative dataset
in this field. As an illustration of the proposed classifier, we only considered two types of flowers among 17 category flowers,
i.e., snowdrop and daffodil flowers. Figure \ref{Eflowers} presented some examples for snowdrop and daffodil flowers.

%\begin{figure}[H]
%\centering
%\scalebox{0.18}[0.2]{\includegraphics{RealData1fig.png}}\vspace{-3mm}
%\caption{\footnotesize{Means of six metrics for eight classifiers in the flowers image data. }}
%\label{flowerresults}
%\end{figure}

To analyze the aforementioned flower image data via the proposed classifier and some competing classifiers
such as the BTLR, KNN, RF, C-SVC, $\nu$-SVR, L-SVR and CNN similar to those given in Section \ref{Simulation},
we require extracting the image data associated with snowdrop and daffodil flowers for 160 flower images, i.e., $N=160$.
To this end, we rescale each of 160 images as $50 \times 60\times 3$ tensor (i.e., $\bmX_i$ with $(I_1,I_2,I_3)=(50,60,3)$), 
where the $50\times 60$ is the size of an image in pixels and the value $3$ represents the cells needed to store information about RGB color.
%	the values $50, 60$ and $3$ represents the height (vertical spatial dimension) of the image in pixels, the width (horizontal spatial dimension) of the image in pixels, and the color channels (RGB: red, green, blue), respectively. 	
We randomly divide $80\%$ of 160 images into training data (i.e., $\bmX_i$ for $i=1,\ldots,128$),
which are used to estimate tensor $\bmW$ in the considered logistic tensor regression model (\ref{equ1}),
and 20\% of 160 images into testing data (i.e., 32 images), which are used to assess the performance of the proposed classifier and
the competing classifiers in terms of the sensitivity, specificity, AUC, accuracy, precision and F1 score
defined in Section \ref{Simulation}. Based on each of 100 replications for the aforementioned random repartition of 160 images,
we utilize the aforementioned variational Bayesian method with the hyperparameters given in Section 3.2
to estimate $\bmW$ and predict the classes of flowers via the proposed classifier and the competing classifiers.
For 100 classification prediction results of flowers and their true classes,
we calculate means and standard deviations of six metrics for each of eight classifiers.
Results with Max-ELBO $R$ value are shown
in Table \ref{tab:performance_comparison_for_real_data}.
% and Figure \ref{flowerresults}.

From Table \ref{tab:performance_comparison_for_real_data},
% and Figure \ref{flowerresults}, 
it is easily seen that the proposed classifier
outperforms other classifiers in that the former has the largest specificity, AUC, accuracy, precision and F1 score among eight classifiers,
and its sensitivity value is larger than $90\%$.
Although the L-SVR classifier has the largest sensitivity among eight classifier and a relatively large AUC (i.e., 95.6\%),
its specificity, accuracy, precision and F1 score are the smallest among eight classifiers,
which implies that the L-SVR classifier cannot identify true negatives well.
Compared with other classifiers, the proposed classifier could improve the specificity, AUC, accuracy, precision and F1 score,
%by more than $3.95\%$, $0.32\%$, $2.25\%$, $3.00\%$ and $2.04\%$, respectively,
and had the relatively large sensitivity (i.e., 90.90\%), specificity (i.e., 92.39\%) and AUC (i.e., 97.52\%).
The above findings indicate that the proposed classifier behaves better than the competing classifiers in terms of six metrics.

\begin{table}[H]	
	%	\centering
	\fontsize{7}{8}\selectfont
	\begin{center}
		\caption{Means (standard deviations) of six metrics (\%) for eight classifiers via 100 replications in the flowers image data.}
		\label{tab:performance_comparison_for_real_data}
		%				\vspace{0.2cm}
		\renewcommand{\arraystretch}{1.3} \tabcolsep 0.10in %\doublerulesep 1.5pt   NN
		\begin{tabular}{ccccccc} \hline\hline
			Method&Sensitivity&Specificity&AUC&Accuracy&Precision&F1-score \\ 			\hline
			VBLTR&90.90(6.93)&{\bf 92.39}(8.04)&{\bf 97.52}(2.50)&{\bf 91.69}(5.32)&{\bf 92.77}(6.60)&{\bf 91.60}(5.19) \\
			%					VBLTR&90.80(0.10)	 &{\bf 91.18}(0.10)	 &96.99(0.05)	 &{\bf 91.07}(0.07)	 &{\bf 91.73}(0.08)	 &{\bf 90.89}(0.08)	\\
			BTLR&  87.63(0.10)	 &83.73(0.12)	 &90.96(0.08)	 &85.63(0.08)	 &84.60(0.11)	 &85.61(0.09)	\\
			C-SVC&89.63(7.78)&88.74(10.98)&96.57(2.80)&89.34(5.79)&89.99(8.91)&89.33(5.52)\\ 	
			$\nu$-SVR&91.61(6.75)&87.42(10.31)&97.21(2.51)&89.63(5.49)&88.52(7.63)&89.73(5.05)\\
			L-SVR&{\bf 99.54}(1.73)&22.11(10.92)&95.60(3.08)&60.53(8.10)&55.79(8.36)&71.14(6.92)\\
			KNN&98.80(2.56)&49.57(12.02)&93.74(4.40)&73.75(7.78)&65.82(9.57)&78.58(6.96)\\
			RF&82.37(9.97)&83.30(11.72)&89.83(5.45)&82.78(6.55)&83.71(10.11)&82.34(6.81)\\
			CNN&80.98(20.75)&79.00(19.63)&86.65(16.74)&79.88(14.45)&80.21(17.64)&78.79(18.10)\\
			\hline\hline
		\end{tabular}
	\end{center}
\end{table}

%\begin{table}[H]	
%%	\centering
%\fontsize{7}{8}\selectfont
%\begin{center}
%	\caption{Means (standard deviations) of six metrics (\%) for eight classifiers via 100 replications in the Chest X-Ray imaging data.}
%	\label{tab:performance_comparison_for_real_data2}
%	\vspace{0.2cm}
%	\renewcommand{\arraystretch}{1.5} \tabcolsep 0.10in %\doublerulesep 1.5pt   NN
%	\begin{tabular}{ccccccc} \hline\hline
%		Method&Sensitivity&Specificity&AUC&Accuracy&Precision&F1-score \\ 			\hline
%		VBLTR&93.43(1.35)&{\bf 90.67}(1.75)&{\bf 97.65}(0.33)&{\bf 92.69}(0.82)&{\bf 96.44}(0.68)&{\bf 94.90}(0.62) \\
%		%					dddd&90.22(0.02)	 &88.96(0.02)	 &96.11(0.00)	 &89.88(0.01)	 &95.68(0.01)	 &92.86(0.01)	\\
%		BTLR&81.40(0.15)	 &82.19(0.11)	 &87.78(0.13)	 &81.63(0.14)	 &91.97(0.06)	 &85.95(0.12)	\\
%		C-SVC&88.96(2.13)&89.71(2.61)&96.52(0.48)&89.17(1.11)&95.92(0.94)&92.29(0.89)\\ 	
%		$\nu$-SVR&89.70(3.04)&90.28(3.58)&96.99(0.43)&89.86(1.40)&96.20(1.23)&92.79(1.14)\\
%		L-SVR&91.97(1.88)&89.64(2.34)&96.39(0.58)&91.34(1.09)&96.01(0.87)&93.93(0.84) \\
%		KNN&{\bf 97.75}(0.48)&77.69(2.11)&96.12(0.63)&92.33(0.72)&92.21(0.89)&{\bf 94.90}(0.50)\\
%		RF&95.73(0.75)&81.02(2.04)&95.77(0.66)&91.76(0.72)&93.17(0.83)&94.43(0.52)  \\
%		CNN&92.84(1.41)&87.79(2.19)&96.07(1.02)&91.48(1.11)&95.36(0.86)&94.08(0.83)   \\
%		\hline\hline
%	\end{tabular}
%\end{center}
%\end{table}

%%%%%%%%%%%%%%%%%%%%%%%%RealData实验end%%%%%%%%%%%%%%%%%%%%%%%%%%%%%

\section{Conclusions}

This paper proposes a logistic tensor regression model with tensor covariates and tensor coefficient for retaining the correlation and
structural information in image data, which avoids the defect of the vectorization of image data that may corrupt data intrinsic structure.
The CP decomposition is adopted to reduce the dimensionality of tensor coefficient.
Based on different matrixization forms of tensor decomposition,
this paper constructs different expressions of logistic tensor regression.
To make Bayesian inference and identify sparse structure of tensor coefficient,
this paper considers Bayesian analogue of Lasso and adaptive Lasso together with
the multiway shrinkage priors for marginal factor vectors of tensor coefficient.
A variational Bayesian approach is developed to reduce the computational burden of the traditional MCMC method
in estimating high-dimensional parameters in tensor coefficient and identifying sparsity of tensor coefficient
based on the mean-field assumption. Due to the logistic function involved, it is rather difficult to obtain the closed-form of the ELBO in the variational Bayesian framework
based on the Gaussian prior assumption of marginal factors. To this end, this paper adopts the method of \cite{Jaakkola2000BayesianPE}
to approximate the lower bound of the joint density function of marginal factors and latent variables, which leads to the conjugate priors of marginal factors.
The variational densities for marginal factors deduced from the mean-field assumption are not optimal.
To overcome the difficult, this paper employs the coordinate ascent algorithm to iteratively optimize the variational lower bound.
For classification prediction, this paper presents a predictive density approximation based on variational posterior densities.
The proposed method can be utilized to deal with the image recognition problem and other classification problems associated with tensor covariates. Simulation studies show that the proposed method behaves better than existing classification methods.  Two real examples taken from flower image recognition and chest X-ray image classification are illustrated to demonstrate the effectiveness of the proposed methodologies.

Note that the proposed variational Bayesian method focuses on the mean-field assumption of the joint variational density. That is, the variational density can be factorized across the components of parameters and latent variables. This assumption may be unreasonable in some complicated tensor regression models. In these cases, one can consider conditional variational Bayesian inference to overcome the potential drawbacks of the mean-field based variational inference.

%%%%%%%%%%%%%%%%%%%%%%%%%%%%%%%%%%%%%%%%%%%%%%
%% Acknowledgements                         %%
%% should be provided in the                %%
%% Acknowledgements section.                %%
%%%%%%%%%%%%%%%%%%%%%%%%%%%%%%%%%%%%%%%%%%%%%%
\begin{acks}[Acknowledgments]
	The authors thank the editor, the associate editor, the Editor-in-Chief, and referees for insightful comments that led to important improvements of this manuscript, and Mendeley Data for making the data available.
\end{acks}

%%%%%%%%%%%%%%%%%%%%%%%%%%%%%%%%%%%%%%%%%%%%%%
%% Funding information, if any,             %%
%% should be provided in the                %%
%% funding section.                         %%
%%%%%%%%%%%%%%%%%%%%%%%%%%%%%%%%%%%%%%%%%%%%%%
\begin{funding}
This work was supported by grants from the National Key R\&D Program of China (No. 102022YFA1003701),
the National Natural Science Foundation of China (No. 12271472, 12231017, 12001479, 11871420).
\end{funding}

%%%%%%%%%%%%%%%%%%%%%%%%%%%%%%%%%%%%%%%%%%%%%%
%% Supplementary Material, including data   %%
%% sets and code, should be provided in     %%
%% {supplement} environment with title      %%
%% and short description. It cannot be      %%
%% available exclusively as external link.  %%
%% All Supplementary Material must be       %%
%% available to the reader on Project       %%
%% Euclid with the published article.       %%
%%%%%%%%%%%%%%%%%%%%%%%%%%%%%%%%%%%%%%%%%%%%%%

\begin{supplement}
\stitle{Supplement to ``Variational Bayesian Logistic Tensor Regression with Application to Image Recognition'' (DOI: )}
\sdescription{\\
A. Introduces the details of the derivation of the optimal variational densities for parameters $\bthe_s$'s ($s=1,\ldots,S$).\\
B. Introduces the $\xi$-transformation-based ELBO of the ELBO.\\
C. Introduces the details of the estimation of variational parameters.\\
D. Introduces that the relationship of the proposed method based on $\xi$-transformation and Ploya-Gamma data augmentation approach in term of variational posterior. \\
E. Shows additional simulation studies to investigate the robustness of the proposed method. \\
F. Shows an order-2 tensor data analysis based on chest X-ray image.}
\end{supplement}

%\begin{supplement}
%\stitle{A. Variational Posterior of Parameters.}
%\sdescription{Introduces the details of the derivation of the optimal variational densities for parameters $\bthe_s$'s ($s=1,\ldots,S$).}
%\stitle{A. Variational Posterior of Parameters.}
%\sdescription{Introduces the details of the derivation of the optimal variational densities for parameters $\bthe_s$'s ($s=1,\ldots,S$).}
%\end{supplement}

\vskip 0.2in

\bibliographystyle{ba}
\bibliography{sample}

\end{document}